\documentclass{emulateapj}

\slugcomment{Accepted by ApJ}

\shorttitle{Gas infall towards Sgr~A*}
\shortauthors{Montero-Casta{\~n}o, Herrnstein and Ho}

\begin{document}

\title{Gas infall towards Sgr~A* from the clumpy circumnuclear disk}

\author{Mar{\'{\i}}a Montero-Casta{\~n}o\altaffilmark{1,2,3}, Robeson M. Herrnstein\altaffilmark{4} and Paul T.P. Ho\altaffilmark{1,3}}

\altaffiltext{1}{Harvard-Smithsonian Center for Astrophysics, 60 Garden Street, Cambridge, MA 02138; mmontero@cfa.harvard.edu, pho@cfa.harvard.edu.}
\altaffiltext{2}{Departamento de Astrof{\'{\i}}sica, Facultad de Ciencias F{\'{\i}}sicas, Universidad Complutense de Madrid, 28040-Madrid, Spain.}
\altaffiltext{3}{Institute of Astronomy and Astrophysics, Academia Sinica, P.O. Box 23141, Taipei 106, Taiwan.}
\altaffiltext{4}{Department of Astronomy, Columbia University, 550 West 120th Street, New York, NY 10027; herrnstein@astro.columbia.edu}
\altaffiltext{5}{The Submillimeter Array is a joint project between the Smithsonian Astrophysical Observatory and the Academia Sinica Institute of Astronomy and Astrophysics, and is funded by the Smithsonian Institution and the Academia Sinica.}


\begin{abstract}

We present the first large-scale mosaic performed with the Submillimeter Array (SMA$^{5}$) in the Galactic center. We have produced a 25-pointing mosaic, covering a $\sim$~2$\arcmin$~$\times$~2$\arcmin$ area around Sgr~A*. We have detected emission from two high-density molecular tracers, HCN(4-3) and CS(7-6), the latter never before reported in this region. The data have an angular resolution of 4.6$\arcsec$~$\times$~3.1$\arcsec$, and the spectral window coverage is from -180~km~s$^{-1}$ to 1490~km~s$^{-1}$ for HCN(4-3) and from -1605~km~s$^{-1}$ to 129~km~s$^{-1}$ for CS(7-6).

Both molecular tracers present a very clumpy distribution along the circumnuclear disk (CND), and are detected with a high signal-to-noise ratio in the southern part of the CND, while they are weaker towards the northern part. Assuming that the clumps are as close to the Galactic center as their projected distances, they are still dense enough to be gravitationally stable against the tidal shear produced by the supermassive black hole. Therefore, the CND is a non-transient structure.

This geometrical distribution of both tracers suggests that the southern part of the CND is denser than the northern part. Also, by comparing the HCN(4-3) results with HCN(1-0) results we can see that the northern and the southern parts of the CND have different excitation levels, with the southern part warmer than the northern.

Finally, we compare our results with those obtained with the detection of NH$_3$, which traces the warmer and less dense material detected in the inner cavity of the CND.  We suggest that we are detecting the origin point where a portion of the CND becomes destabilized and approaches the dynamical center of the Milky Way, possibly being impacted by the {\it southern streamer} and heated on its way inwards.

\end{abstract}




\section{Introduction}

The Galactic center, located at a distance of 8 kpc \citep{rei93} (1$\arcsec$ corresponds to 0.038~pc), is our unique laboratory to study the interactions between a supermassive black hole (4~$\times$~10$^{6}$ M$_\odot$, \cite{ghe05,sch05}) and its surrounding environment. This region is composed of a very dense and warm interstellar medium (ISM), mostly condensed in Giant Molecular Clouds (GMCs). These GMCs have different characteristics than those GMCs found in the Galactic disk. In the Galactic center, the GMCs are warmer, denser, and much more turbulent. The dynamical center of the Milky Way is occupied by a nonthermal compact radio source, the supermassive black hole Sgr~A*. Surrounding this source, three arcs of ionized gas are present (the western arc, the northern arm and the extended bar, \cite{rob93}), forming the minispiral or Sgr~A~West. Both Sgr~A* and the minispiral are surrounded by a ring-like structure, the circumnuclear disk (CND). The CND is composed of a mixture of neutral atomic and molecular gas and dust \citep{gen85,mez89}. The CND has an inner radius of $\sim$2~pc \citep{gat84,gen85,jac93,chr05} and it extends until $\sim$~5~pc \citep{har85,gen85}. It is tilted with respect to the Galactic plane ($\sim$~70$\arcdeg$) and its main motion is rotation around Sgr~A* \citep{gat84,gen85,lug86,gus87}. The inner cavity that the CND encloses is devoid of dust \citep{bec82}, but it is occupied by ionized gas (minispiral) \citep{gat84}, neutral molecular gas \citep{har85,her02} and a large amount of neutral atomic gas \citep{jac93}. Also, inside the cavity, occupying the central parsec, stars in different evolutionary stages have been found \citep{pau06}. The stars are orbiting Sgr~A* 
and they could have been formed in-situ \citep{pau06} or spiraled inwards after formation \citep{gur05}.

\cite{gus87} have previously detected the clumpiness of the CND, calculating a dynamical lifetime of 10$^{4}$~--~10$^{5}$~yr. However, later studies by \cite{jac93}, \cite{shu04} and \cite{chr05} estimate that the density is $\sim$~(3~--~4)~$\times$~10$^{7}$~cm$^{-3}$. The density of the clumps would allow them to overcome the tidal shear from Sgr~A*, increasing the lifetime of the CND to 10$^{7}$~yr.

\subsection{Molecular studies}

The Galactic center has been studied numerous times in different molecular gas tracers. 
\cite{coi99,coi00,mcg01,her02,her05} studied the emission from four different ammonia rotation inversion transitions (NH$_3$(1,1), (2,2), (3,3) and (6,6)) using the Very Large Array (VLA). Ammonia is a moderately high-density ($\sim$~10$^{4}$~cm$^{-3}$), and high-temperature (23 to 412~K, (1,1) to (6,6) transitions) tracer. These studies showed that the colder gas (NH$_3$(1,1) and (2,2)) is highly extended in this region, showing no strong emission in the CND. However, NH$_3$(3,3) (energy above ground $\sim$~125~K) is more concentrated towards the center, proving to be much more useful for tracing the CND, although still not found any closer to the supermassive black hole than the CND. NH$_3$(6,6), on the other hand (tracing material at $\sim$~412~K), is found very close to Sgr~A*, predominantly inside the CND, well within the inner cavity, while it is much fainter further away from the CND. Therefore, very dense and warm molecular gas is located in the inner 1.5~pc of the Galaxy.

Studies of other high-density tracers have also been carried out. In particular, HCN has proved to be a very useful molecular gas tracer in the Galactic center environment, providing important results regarding the structure of the CND. \cite{gus87} studied HCN(1-0) emission using the Hat Creek Interferometer, obtaining a 10$\arcsec$ resolution. The emission from this transition was detected along the CND, with the emission in the southern part stronger than the emission detected in the rest of the CND. Also, a gap was found in the eastern part of the CND. High-resolution results were also achieved by \cite{wri01} using the Berkeley-Illinois-Maryland Array (BIMA). However, \cite{wri01} noted that self-absorption affects HCN(1-0) (and HCO$^{+}$(1-0), which is closely correlated) due to the intervening molecular material along the line of sight to the Galactic center. This problem was also addressed by \cite{chr05}, who published their HCN(1-0) data from the Owens Valley Radio Observatory (OVRO) millimeter interferometer, with a 5$\arcsec$ resolution. In these latest results, the HCN emission is detected along the CND and also in various narrow structures outside of it, such as the linear filament. However, because of the self-absorption problem, the use of higher transitions, which will be less sensitive to the cooler and more diffuse gas along the line of sight, should lead to improvements.

HCN(3-2) emission was detected by \cite{jac93} using the IRAM 30-m telescope and by \cite{mar95} using the 15-m James Clerk Maxwell Telescope (JCMT). In both cases, emission is detected along the CND, but it is remarkably stronger towards the south of the CND. The detection of HCN(4-3) emission (which traces material at $\sim$~25~K, and has a critical density value of $\sim$~10$^{8}$~cm$^{-3}$; \citealt{cho00,tak07}) in the Galactic center region was also reported by \cite{mar95} using the JCMT, achieving a 15$\arcsec$ resolution. The HCN(4-3) emission is also stronger towards the south of the CND, with very weak emission towards the north. Moreover, a dynamical model of a rotating torus developed by this group supported the idea that the northern and the southern parts of the CND are independent structures (each of them with a different inclination, e.g. a warped disk), following a common rotation pattern around Sgr~A*. The self-absorption problem is much less evident for these higher-excitation transitions, consistent with the a priori expectation that absorption is produced by lower excitation material along the line of sight.

While these higher HCN lines are much more reliable tracers for studying the structure of the CND, higher-resolution has not been available until now. The Submillimeter Array (SMA) is the first interferometer equipped with 350~GHz receivers, and our maps image the CND in the higher excitation material.

CS has also been studied in the Galactic center in various transitions ((2-1), (3-2), (5-4) and (7-6) \citealt{ser89,ser92}). However, no previous results on (7-6) have been reported in the CND. Our study images the distribution of CS(7-6) emission (which traces material at $\sim$~49~K and has a critical density of $\sim$~10$^{7}$~cm$^{-3}$, \cite{cho00,tak07}) in the central 4~pc of the Milky Way.

The NH$_3$ studies allow a determination of kinetic temperatures. The HCN and CS studies allow estimates of the H$_2$ densities. Together, they define the high excitation environment of galactic nuclei.


\section{Observations}

\begin{figure}
\begin{center}
\epsscale{1}
\plotone{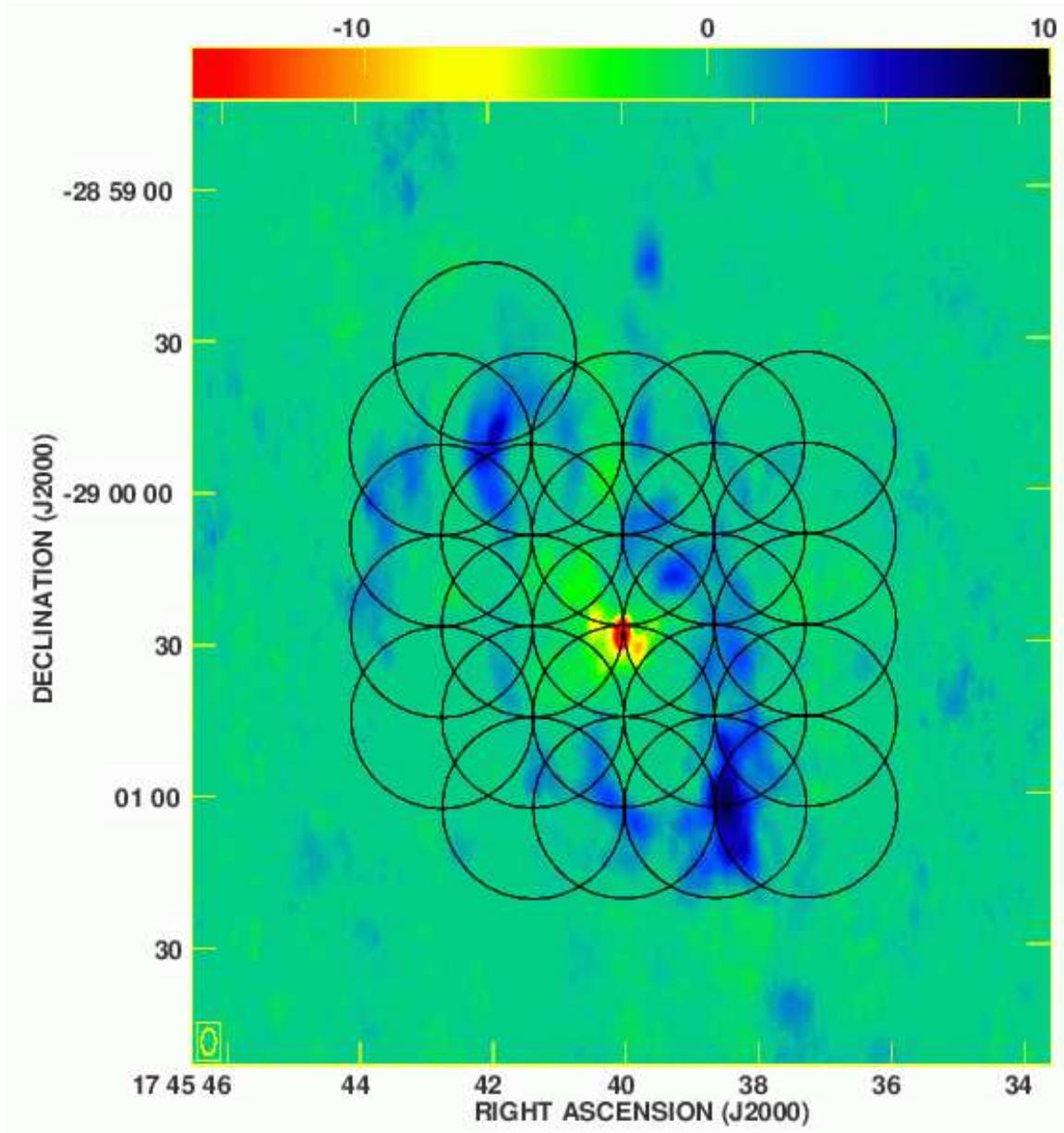}
\end{center}
\caption{25-pointing mosaic. The background image is HCN(1-0) from \cite{chr05}.\label{pointings.fig}}
\end{figure}

Observations of the HCN(4-3) ($\nu$~=~354.5054759 GHz) and CS(7-6) transitions ($\nu$~= 342.8830000 GHz) were made with the Submillimeter Array \citep{ho04} in the compact configuration on 2005 July 1 and 7 and August 13 and in the subcompact configuration on 2007 May 5. We produced a 25-pointing mosaic (figure \ref{pointings.fig}) covering a $\sim$~2$\arcmin$~$\times$~2$\arcmin$ area, which encompasses Sgr~A*, the minispiral and the CND. The central position of the mosaic was at $\alpha_{J2000.0}$~=~17$^h$45$^m$40.00$^s$, $\delta_{J2000.0}$~=~-29$\arcdeg$00$\arcmin$26.60$\arcsec$, and the rest of the pointings were Nyquist sampled at half-beam spacings, effectively increasing the sensitivity of the mosaic (the primary beam for these frequencies is 36$\arcsec$). The data consisted of two 2~GHz bandwidths (upper sideband and lower sideband, USB and LSB, respectively) divided into 24 spectral windows, each of them composed of 128 channels with a velocity resolution of 0.7~km~s$^{-1}$.  The HCN(4-3) was observed in the USB and CS(7-6) in the LSB. The velocity coverage for the USB was from -180~km~s$^{-1}$ to 1490~km~s$^{-1}$, centered on the HCN(4-3) line at {\it v}$_{LSR}$~=~0~km~s$^{-1}$ and from -1605~km~s$^{-1}$ to 129~km~s$^{-1}$ for the LSB, centered on the CS(7-6) line at {\it v}$_{LSR}$~=~0~km~s$^{-1}$. We averaged every 5~km~s$^{-1}$ since molecular lines in the vicinity of Sgr~A* are very broad (FWHM~$\approx$~100~km~s$^{-1}$; \citealt{har85}). Therefore the degraded velocity resolution was adequate to resolve the lines. The continuum subtraction was performed in the uv plane. The total integration time per pointing was 30 minutes, but because of the distribution of the pointings, the actual time for the central 1.5$\arcmin$ covering the CND was 90 minutes. 
The phase calibrator was 1733-130 (NRAO~530), with a flux value of 0.84~Jy for the first two tracks and 0.79~Jy for the third and fourth tracks (the fourth track in the subcompact array). Also, during the fourth track we used 1751+096 as a phase calibrator as well, with a flux value of 1.75~Jy. The fluxes were determined by the SMA project in the immediate period around the experiment.

The data were calibrated using the MIR software package developed for the Owens Valley Radio Observatory (OVRO), which was modified for the SMA, and the imaging process was done using both the Multichannel Image Reconstruction Image Analysis and Display (MIRIAD) and the NRAO Astronomical Imaging Processing System (AIPS). To produce the dirty map we used the task invert in MIRIAD, with the systemp option to decrease the weight for visibilities with abnormally high system temperatures. Afterwards, we subtracted the continuum using line-free channels from both sides of the line for the HCN(4-3), and only one side for the CS(7-6), since the latter line was located close to the edge of the passband in order to be able to observe both molecular lines simultaneously. We cleaned the map using the task mossdi2 in MIRIAD, cleaning to a 1.5~$\sigma$ level. The integrated intensity maps were produced using the MOMNT task in AIPS, with a minimum flux cutoff of 0.5~Jy and 0.3~Jy for the HCN(4-3) data and the CS(7-6) data, respectively. The flux cutoff was chosen based on the noise threshold, to suppress noise contribution, since for a simple summation the noise will add and narrow line features will become effectively filter diluted. Natural weighting of the uv data produced an image with a synthesized beam of 4.6$\arcsec$~$\times$~3.0$\arcsec$ with a position angle of 4.1$\arcdeg$ for the HCN(4-3) emission and 4.5$\arcsec$~$\times$~3.1$\arcsec$ and a position angle of -4.4$\arcdeg$ for CS(7-6). The final RMS per channel is 0.3~Jy~beam$^{-1}$ and the overall achieved RMS sensitivity for the integrated map is 8.4~Jy~beam$^{-1}$~km~s$^{-1}$ (for HCN(4-3)) and 0.2~Jy~beam$^{-1}$ and 3.3~Jy~beam$^{-1}$~km~s$^{-1}$ (for CS(7-6), respectively).

The compact-north (for low declination sources) configuration of the SMA provides a maximum projected baseline distance of 91~m and a minimum of 11~m, while the subcompact configuration has maximum projected baseline of 25~m and a minimum of 6~m. The combination of data from both configurations improves the detection of the extended emission as the inner uv plane is better sampled, while providing a better resolution because of the longest baselines. The data obtained using only the compact configuration show a better resolution, but the negative bowls due to the missing extended emission are quite severe. Eventually, the ultimate goal would be to combine these data with single-dish data in order to further suppress the missing flux problems which are still evident in our maps.


\section{HCN(4-3)}

\begin{figure}
\begin{center}
\epsscale{1}
\plotone{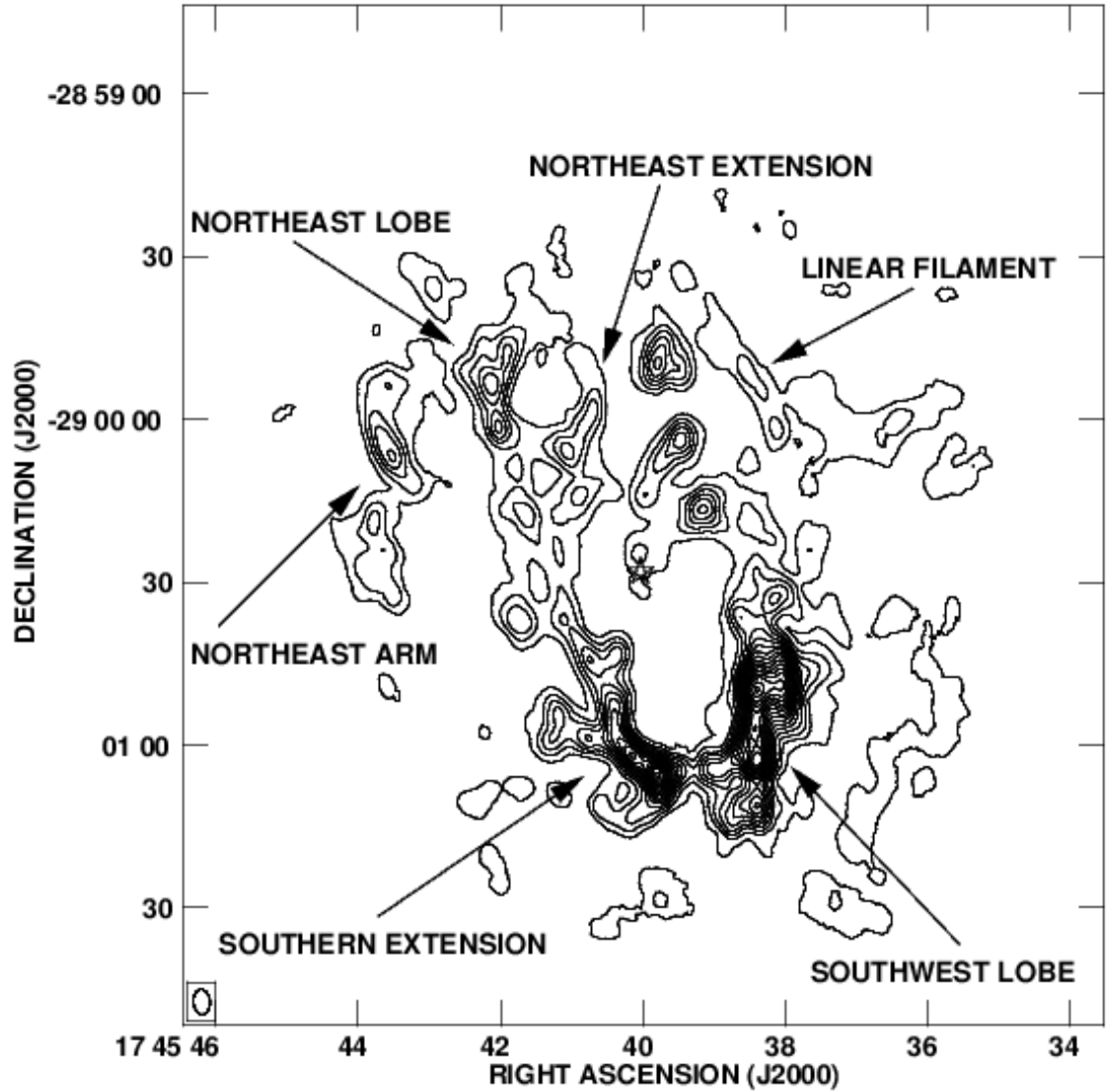}
\end{center}
\caption{HCN(4-3) integrated intensity map. The contour levels are in steps of 6$\sigma$, from 3 to 93$\sigma$, but for the highest contour level, at 97$\sigma$ (2.5~$\times$~10$^{1}$ to 81.7~$\times$~10$^{1}$~Jy~beam$^{-1}$~km~s$^{-1}$). Sgr~A* is marked with a star.\label{hcn4_3.fig}}
\end{figure}

We detected the extended emission of HCN(4-3) within 2~pc of Sgr~A* (figure \ref{hcn4_3.fig}). The southern part of the CND is clearly detected, tracing the southwest lobe and the southern extension (using the same nomenclature as \cite{chr05}), while the emission from the northern part is more sparse and distributed.

\begin{figure}
\begin{center}
\epsscale{1}
\plotone{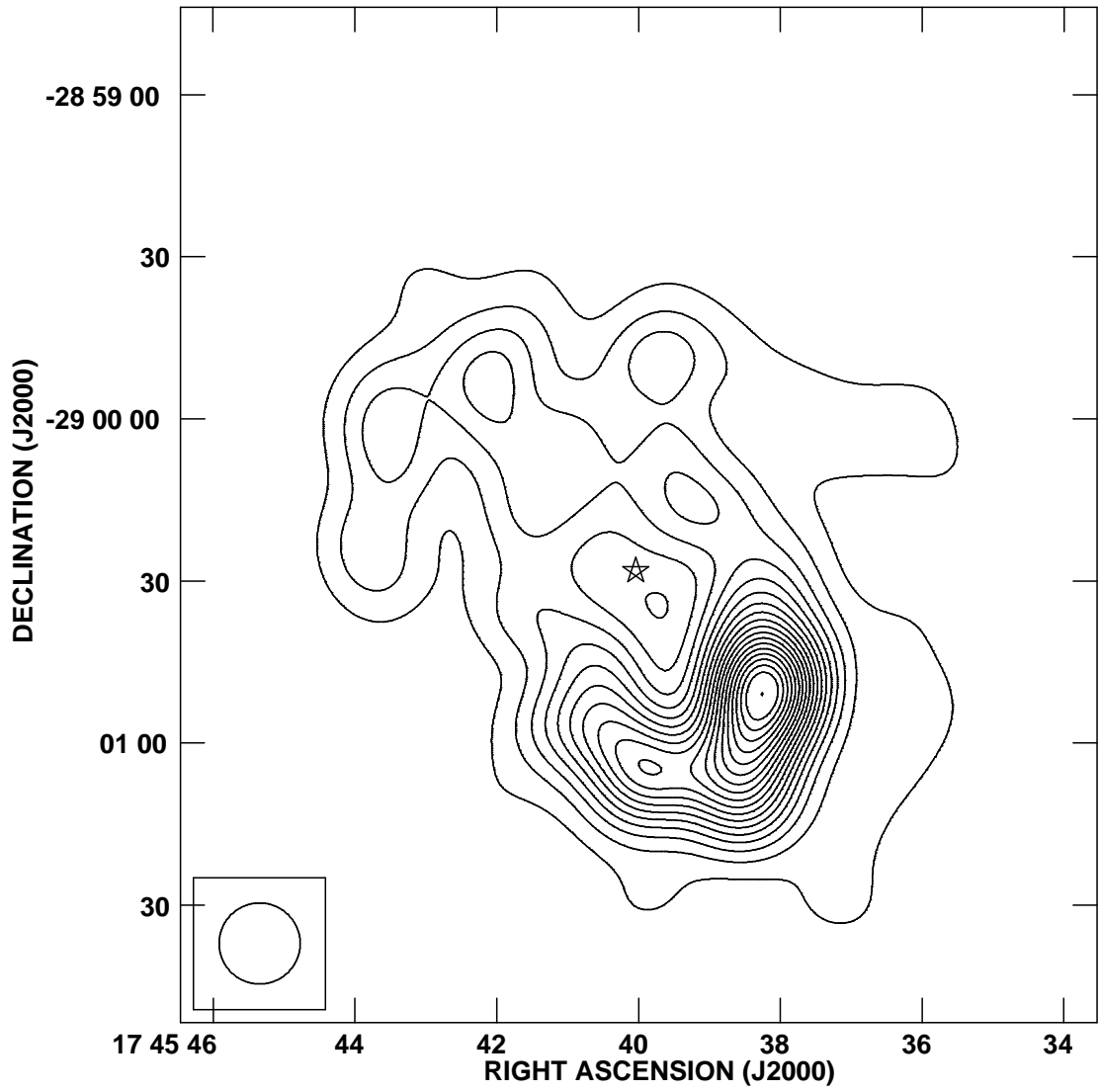}
\end{center}
\caption{HCN(4-3) integrated intensity map smoothed to a 15$\arcsec$ resolution. The contour levels are in steps of 10$\sigma$, from 10 to 180$\sigma$, but the highest contour level, which is 194$\sigma$ (2.7~$\times$~10$^{1}$ to 5.2~$\times$~10$^{3}$~Jy~beam$^{-1}$~km~s$^{-1}$). Sgr~A* is marked with a star.\label{smooth.fig}}
\end{figure}

\cite{mar95} observed a similar distribution of HCN(4-3) emission using the James Clerk Maxwell Telescope (JCMT) (15$\arcsec$ spatial resolution), detecting the strongest emission to the south. Due to our higher resolution image, we can directly determine that the emission is very clumpy around Sgr~A*, i.e., the CND is composed of an array of blobs, in a necklace-like manner, especially towards the north, as was first noted by \cite{gus87}. Furthermore, the clumpy appearance is highly irregular, in terms of size scale, intensity, spacing and distance to Sgr~A*. We compare the HCN(4-3) flux density detected with the SMA to the single-dish flux density detected with the JCMT. After smoothing the SMA resolution to match that of the JCMT, we find that the interferometer detects 61\% of the emission detected with the single-dish over an area that covers 10 JCMT beams. The missing flux density would be spread over many SMA synthesized beams, which are approximately 16 times smaller in area than the JCMT beam (i.e. 161 SMA synthesized beams). The SMA data smoothed to the JCMT resolution (figure \ref{smooth.fig}), however, show a very striking similarity to the JCMT emission map (figure 4 in \citealp{mar95}). This result suggests that the extended emission is a very low level contribution to the final result (21~Jy~km~s$^{-1}$ per SMA synthesized beam, which is at the level of the first contour on figure \ref{hcn4_3.fig}, 3$\sigma$), and most importantly that both the smoothed SMA and the JCMT emission maps are dominated by the clumps.

\begin{figure}
\begin{center}
\epsscale{1}
\plotone{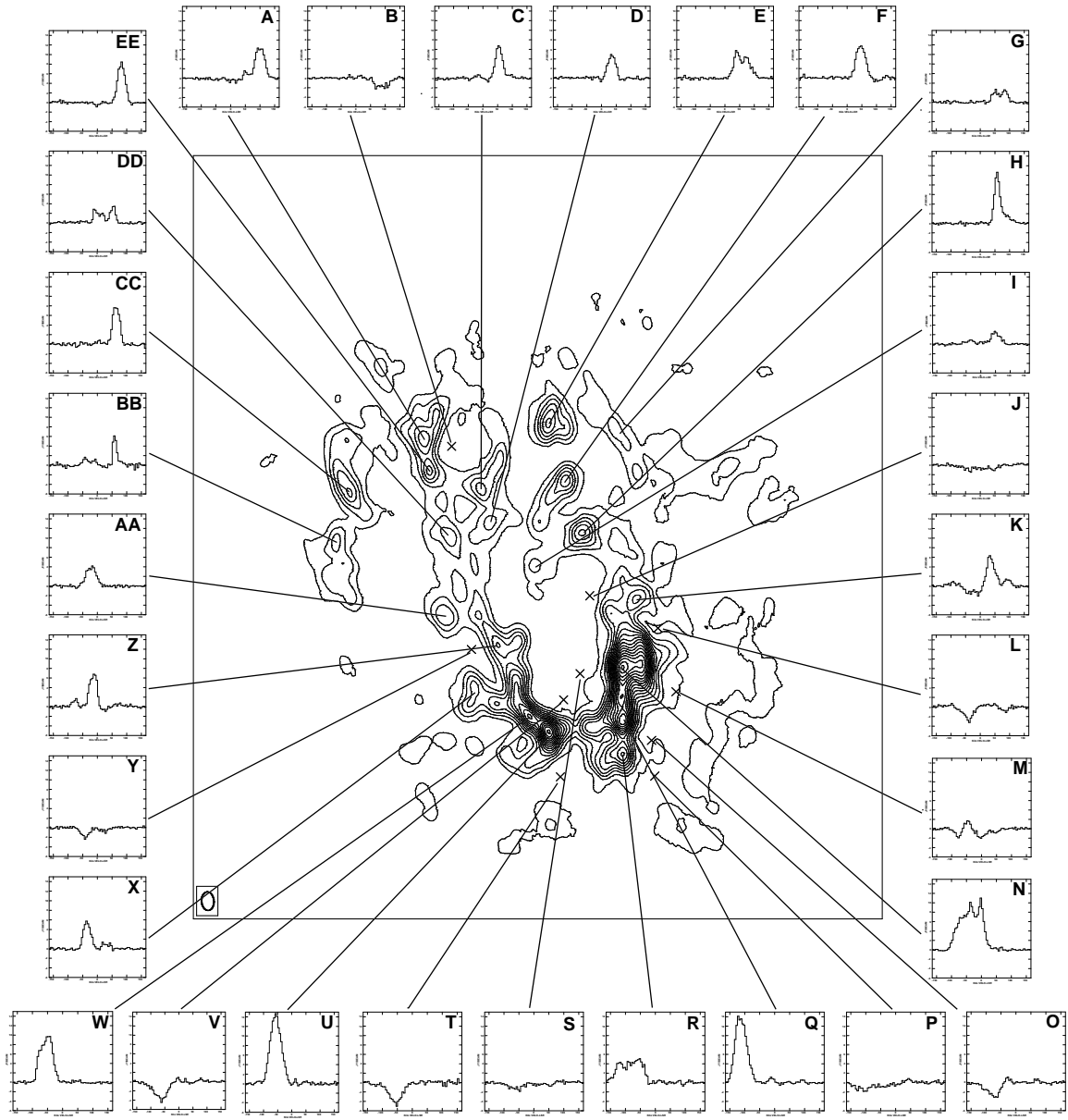}
\end{center}
\caption{HCN(4-3) spectra measured at the positions of the different clumps.\label{spectra_hcn43.fig}}
\end{figure}

\begin{figure}
\begin{center}
\epsscale{1}
\plotone{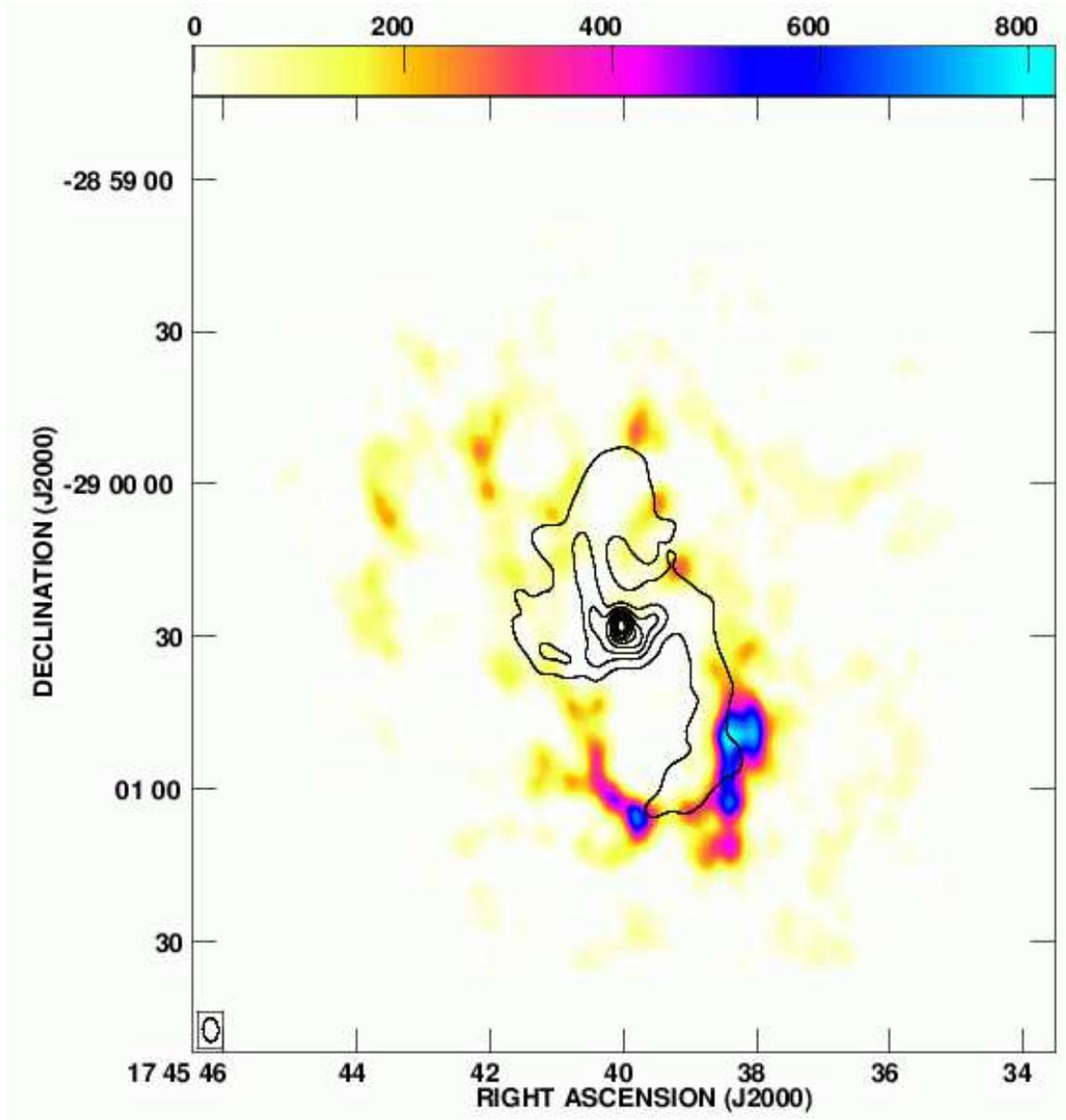}
\end{center}
\caption{6~cm continuum integrated intensity in contours (from \cite{yus87}). HCN(4-3) integrated intensity in false color-scale. Contour levels are in steps of 10\% of the continuum emission peak, from 1~$\times$~10$^{-1}$~Jy~beam$^{-1}$ to 9~$\times$~10$^{-1}$~Jy~beam$^{-1}$. The false color-scale is in ~Jy~beam$^{-1}$~km~s$^{-1}$.\label{hcn43_cont.fig}}
\end{figure}

It is widely accepted that the overall motion of the CND is rotation around Sgr~A*, as we have previously mentioned. We plot the spectra at different peaks (clumps) along the CND to show the kinematics (Fig \ref{spectra_hcn43.fig}). Also, we plot the spectra at different ``empty'' locations where no emission was found close to peculiar-looking spectra to check whether the influence of negative bowls (i.e. the lack of short spacings in our images since extended structures are not recorded) have affected the emission and, therefore, the final result.
We find that spectra A and Q show the largest velocity shifts, with spectrum A redshifted and spectrum Q blueshifted. Spectrum A peaks at $\sim$~105~--~110~km~s$^{-1}$ and spectrum Q at $\sim$~-105~--~-110~km~s$^{-1}$, which agree with the values previously reported by \cite{gus87,jac93,mar95} and \cite{chr05}. The remainder of the features peak at intermediate velocities, in an orderly manner, consistent with a rotation pattern in the CND. With an average radial velocity of 110~km~s$^{-1}$ and a 1.5~pc radius, if we assume that the velocity is purely rotational, the rotation period would be $\sim$~8~$\times$10$^{4}$~yr. We can extract much more information from the study of spectra. In particular, there is a group of features worth observing closely. Firstly, spectrum H is slightly different from the rest. It does not seem to follow the rotation pattern and has a different linewidth, being narrower than the other spectra. These characteristics suggest that this feature may be the ``70~km~s$^{-1}$ cloud'', first reported by \cite{gus87}. It seems to be near the CND, but has a non-negligible radial velocity component ($\sim$~50~km~s$^{-1}$; \citealt{jac93}), much larger than the rest of the clumps, whose predominant motion is rotation. At the same time, spectrum H shows a shock appearance, with a steep blue side and a non-gaussian red tail. Also, clump E shows a remarkably different spectrum, with two velocity components, roughly equal in intensity and an overall broader profile than nearby clumps. Clump E lies to the north of the minispiral northern arm (figure \ref{hcn43_cont.fig}), believed to be interacting with the CND and creating a gap where the ionized gas is infalling towards Sgr~A* \citep{wri01}. In fact, \cite{chr05} consider clump E part of the {\it CND northern arm}. \cite{jac93} reported the detection of neutral gas also associated with the minispiral northern arm as far as 3~pc from the dynamical center of the Milky Way, consistent with a scenario of infalling gas through the CND gap. Clump G, located in close proximity to clump E, also shows a two-velocity component broad profile. Clump G is part of what \cite{chr05} called the {\it linear filament}, a ridge of gas that appears to be connecting the {\it western streamer} and the CND \citep{mcg01}, though it might be a projection effect. On the other hand, we observe a very peculiar (and broad) double-peak spectrum in clump N, which also seems to be in the path of the minispiral (figure \ref{hcn43_cont.fig}). \cite{ser85} reported the western arc of the minispiral to have the same velocity field as the CND, therefore suggesting a connection between the molecular ring and Sgr~A West. However, clump N might be also affected by a different process. When comparing spectrum M with spectrum N, we observe that both show two 'dips' at roughly the same velocities ($\sim$~-60~km~s$^{-1}$, $\sim$~10~km~s$^{-1}$), even though spectrum M was taken a little farther from the region apparently influenced by the minispiral.  Also, spectrum L shows that there are strong negative bowls in the proximity of clump N because of the lack of short-spacings, so that these spectra are probably affected by this problem. We will go back to discuss this spectrum a little later.

Clump DD does show a very clear two-velocity component broad profile, and may be related to the interaction with the minispiral. Cloud DD is located at the end of the extended bar of Sgr~A West (figure \ref{hcn43_cont.fig}). We could be observing a similar behavior as the one detected for clump E. Also, absorption is probably not the answer for this feature due to its uniqueness among the nearby clumps. We would have expected the same profile to be shown by the rest of the adjacent clumps (such as C, D and AA), but that is not the case. We have not found strong negative bowls in the proximity of clump DD. Therefore the lack of short-spacings is probably not affecting the line profile for this clump.

As \cite{jac93} and \cite{mar95} previously remarked, we do detect emission in the eastern and northeastern parts of the CND, a section that is heavily affected by absorption by foreground cold clouds when using a lower J transition line \citep{gus87}. The emission peaks detected in this region (D and DD) are indeed strong, demonstrating that the higher temperature gas in this region can be discerned once the cold gas along the line of sight is suppressed.

Spectra BB and CC show very narrow profiles, especially spectrum BB. They are located outside of the main CND structure, in what we will call the {\it northeast arm}, therefore farther from the supermassive black hole. The greater distance from Sgr~A* implies that the gravitational pull from the center will be less important, and the narrower linewidths may reflect a weaker interaction.

Spectrum K shows a peculiar profile as well. Observing the profile of spectrum L, closely located, we infer that clump K is affected by the lack of short-spacings in the data. A similar behavior seems to be affecting clump Z, easily understood when comparing to spectrum Y.

Clumps U, W and Q are probably also affected by the negative bowls (see spectra T, V and O), which are much stronger in the southern part of the CND, indicating that the loss of extended emission is more significant in that region than in the northern part, where only clump A seems to be affected.

Spectrum R seems to be especially affected by the short-spacings problem, resulting in a very broad but not very strong spectrum. In summary, the HCN(4-3) data  undersample the extended emission. This is especially remarkable in the southern part of the CND. However, such problem does not prevent the very strong detection of this molecular tracer along the CND and even further away, in the linear filament and the {\it northeast arm}.

\begin{figure}
\begin{center}
\epsscale{1}
\plotone{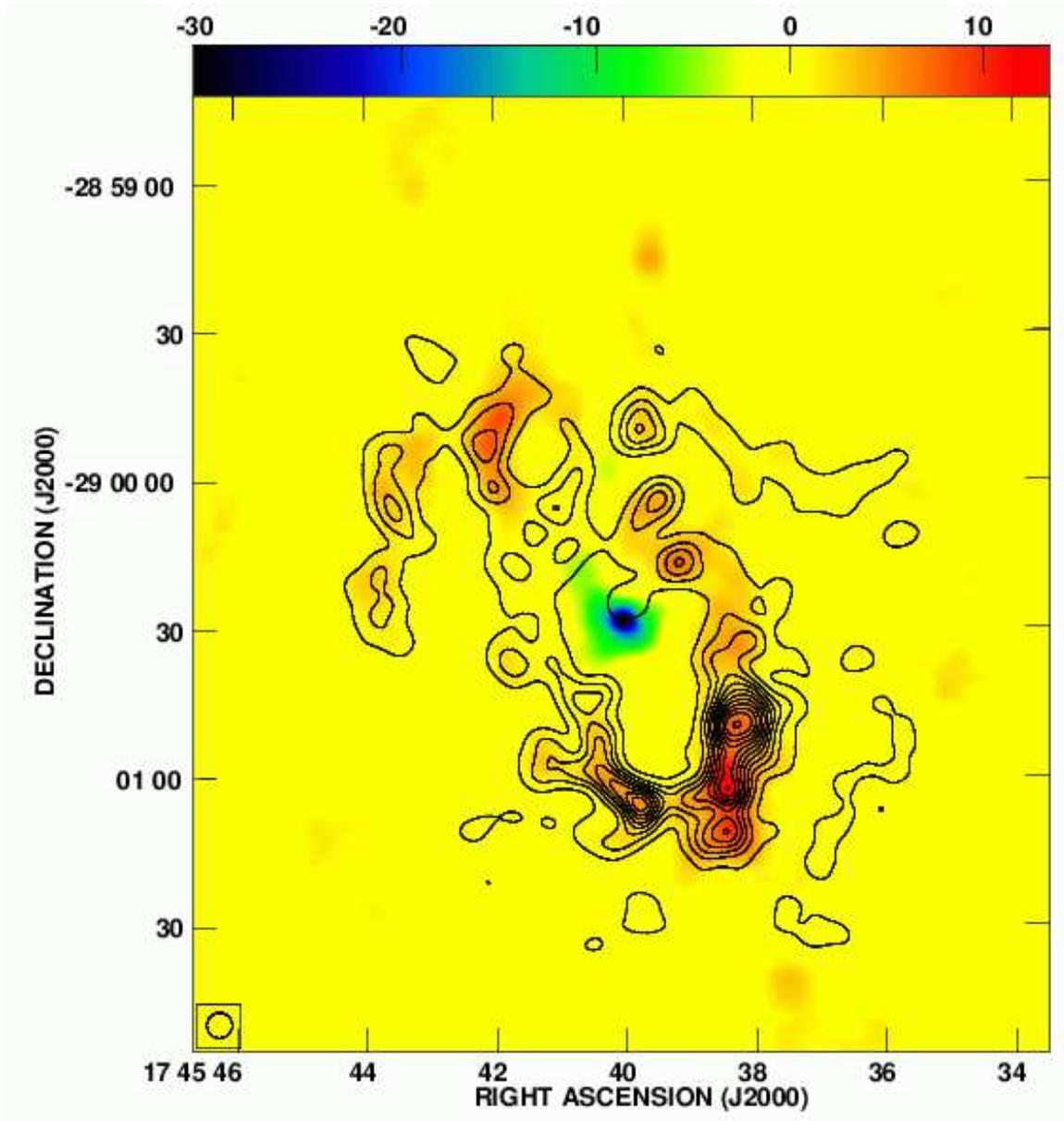}
\end{center}
\caption{HCN(4-3) integrated intensity in contours. HCN(1-0) integrated intensity in false color-scale from \cite{chr05}. Contour levels are in steps of 6$\sigma$, from 3$\sigma$ to 63$\sigma$, except for the highest contour level, at 70$\sigma$ (5.9~$\times$~10$^{1}$ to 13.7~$\times$~10$^{1}$~Jy~beam$^{-1}$~km~s$^{-1}$). The false-color scale is in ~Jy~beam$^{-1}$~km~s$^{-1}$. Sgr~A* is seen in absorption in HCN(1-0).\label{hcn4_3_hcn.fig}}
\end{figure}

Comparing our data with the HCN(1-0) data from the Owens Valley Radio Observatory by \cite{chr05} (figure \ref{hcn4_3_hcn.fig}, both maps convolved to the same resolution and integrated over the same velocity range) we can see that both transitions coincide tracing the southern part of the CND, and even in some parts of the northern lobe of the CND, but HCN(1-0) is stronger to the north as compared to HCN(4-3). This result suggests that the northern and the southern parts of the CND have different excitation conditions, with the southern part warmer than the northern part. The strongest peak in HCN(1-0) in the {\it southwest lobe} does not coincide with the strongest peak in HCN(4-3). It appears that the lower-excitation line is stronger towards the tip of the CND, where the most blueshifted material is detected, while HCN(4-3) is strongest north of that position, in what we have called clump N (figure \ref{spectra_hcn43.fig}). However, HCN(1-0) in the position of clump N is affected by self-absorption \citep{wri01}. Therefore, the difference in gas distribution could be due to this effect.

\begin{figure}
\begin{center}
\epsscale{0.3}

\plotone{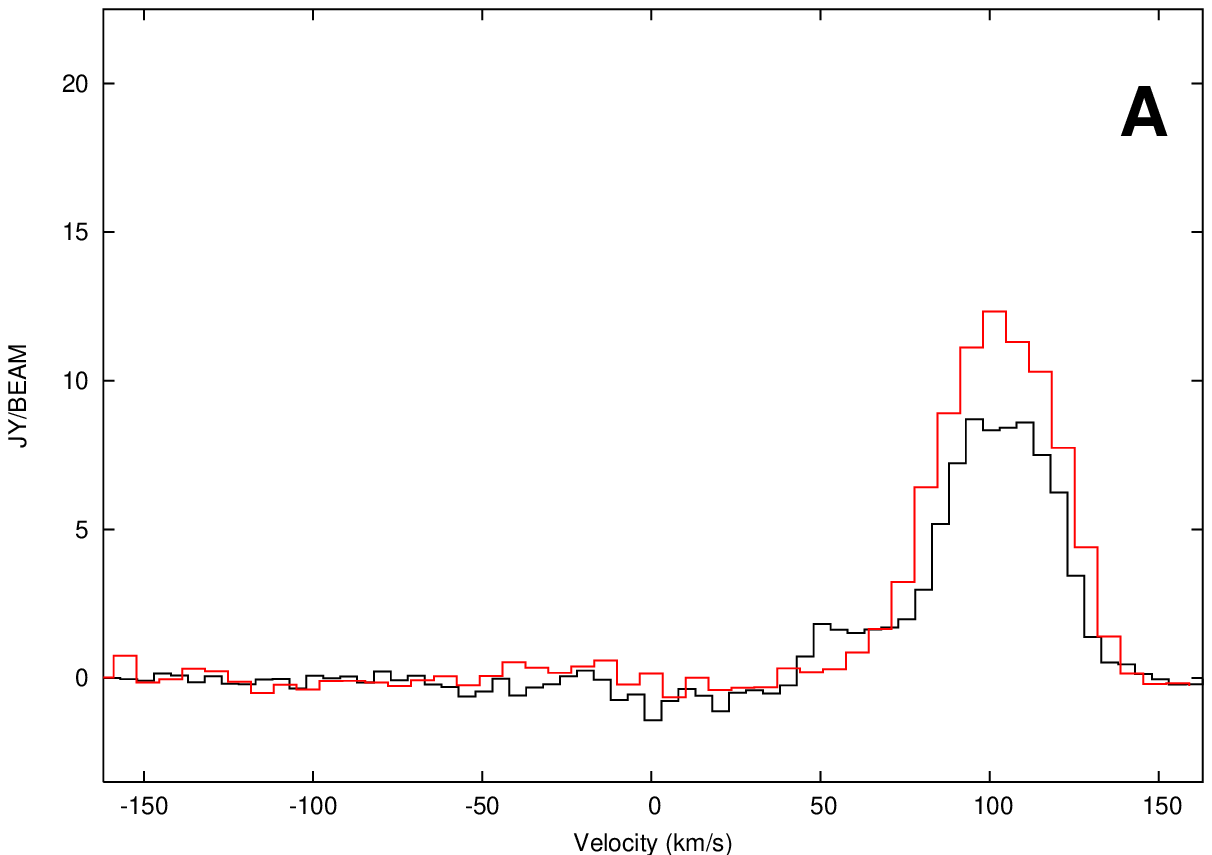}
\plotone{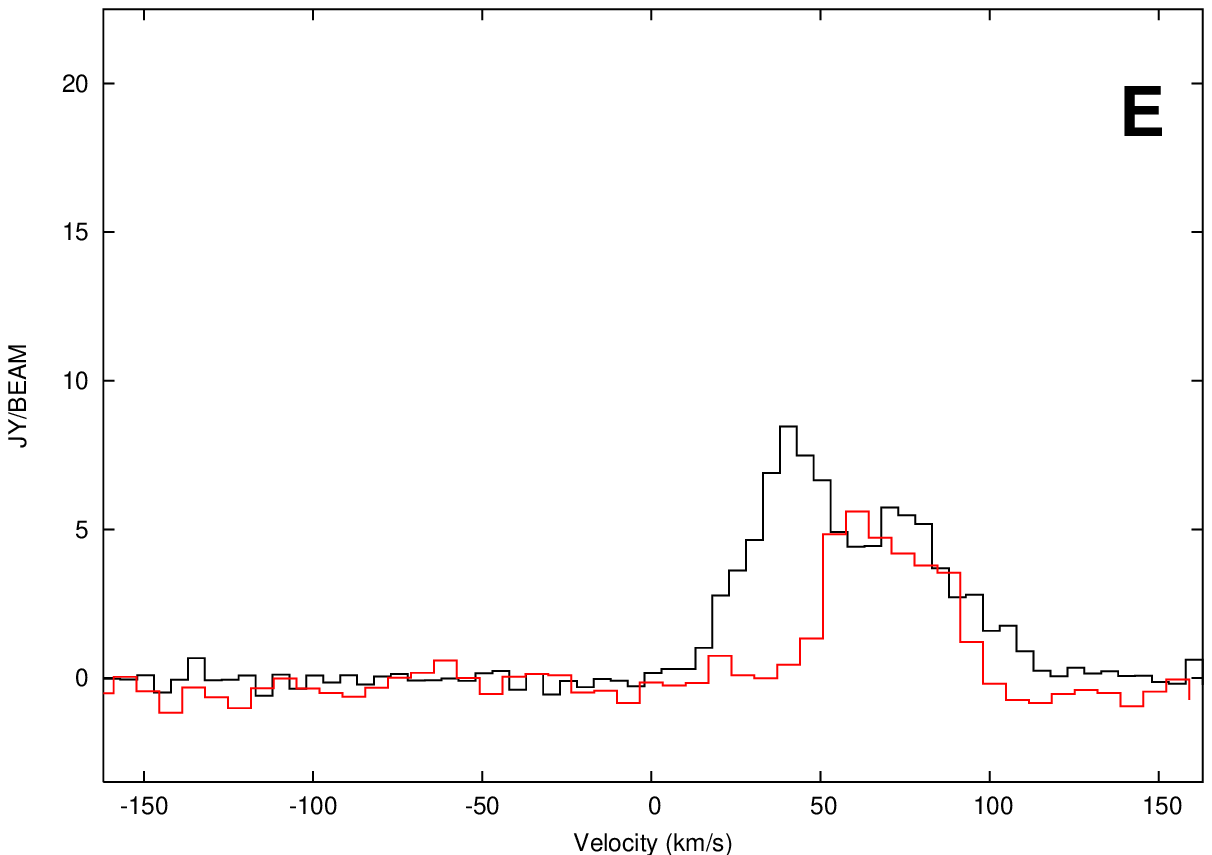}
\plotone{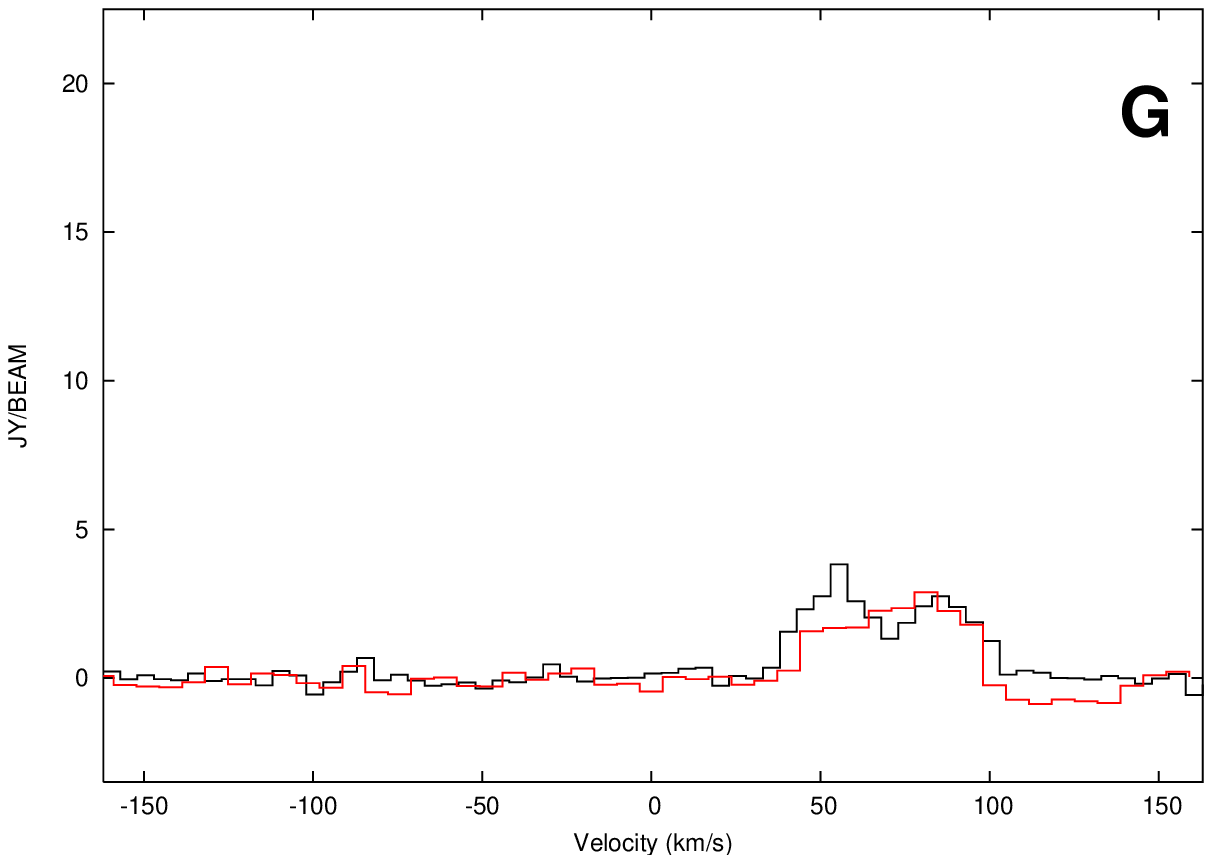}

\plotone{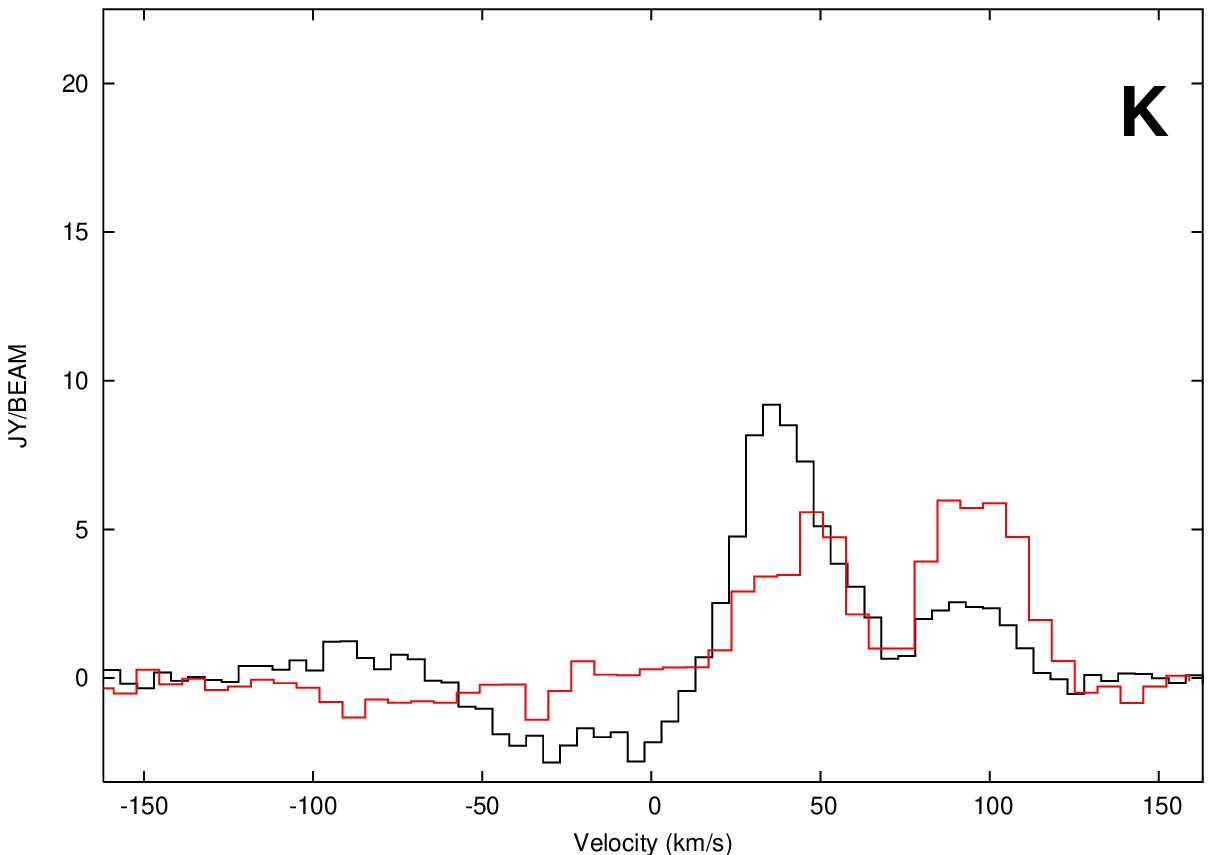}
\plotone{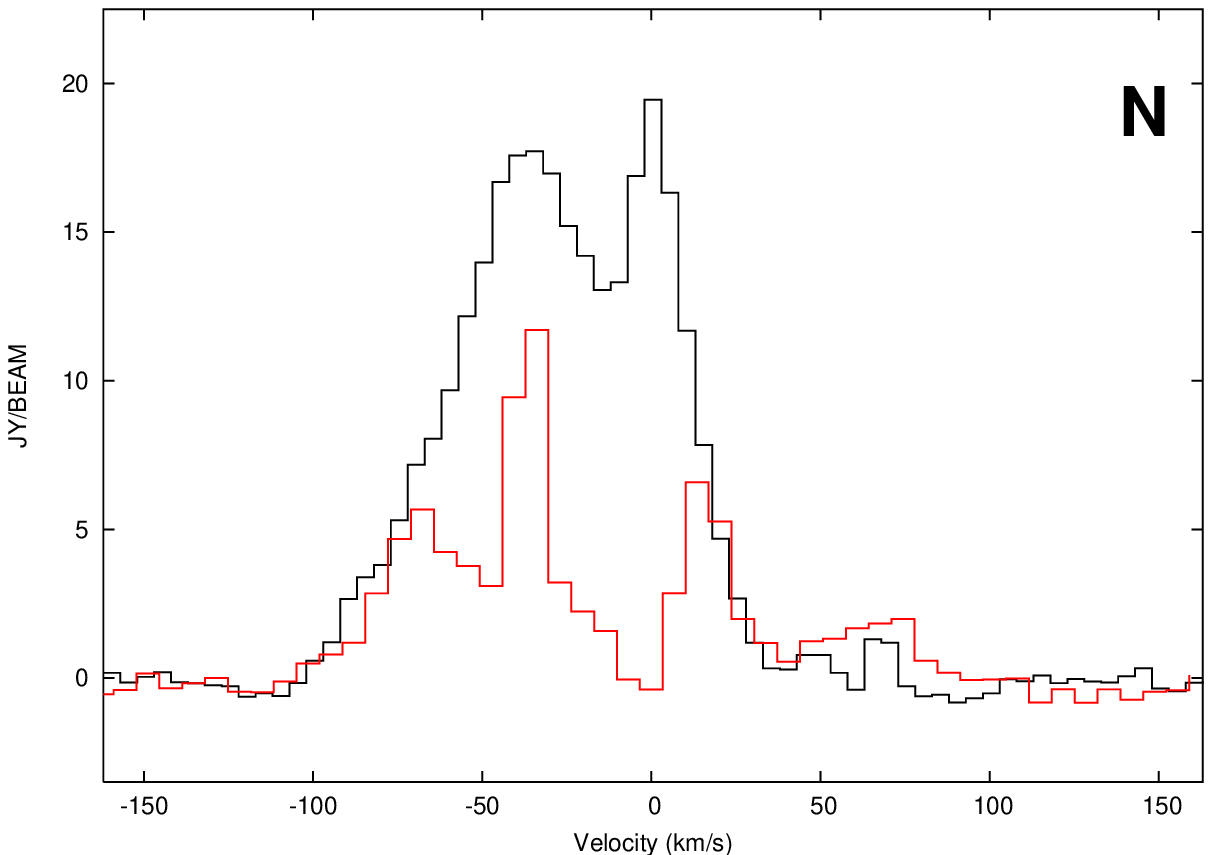}
\plotone{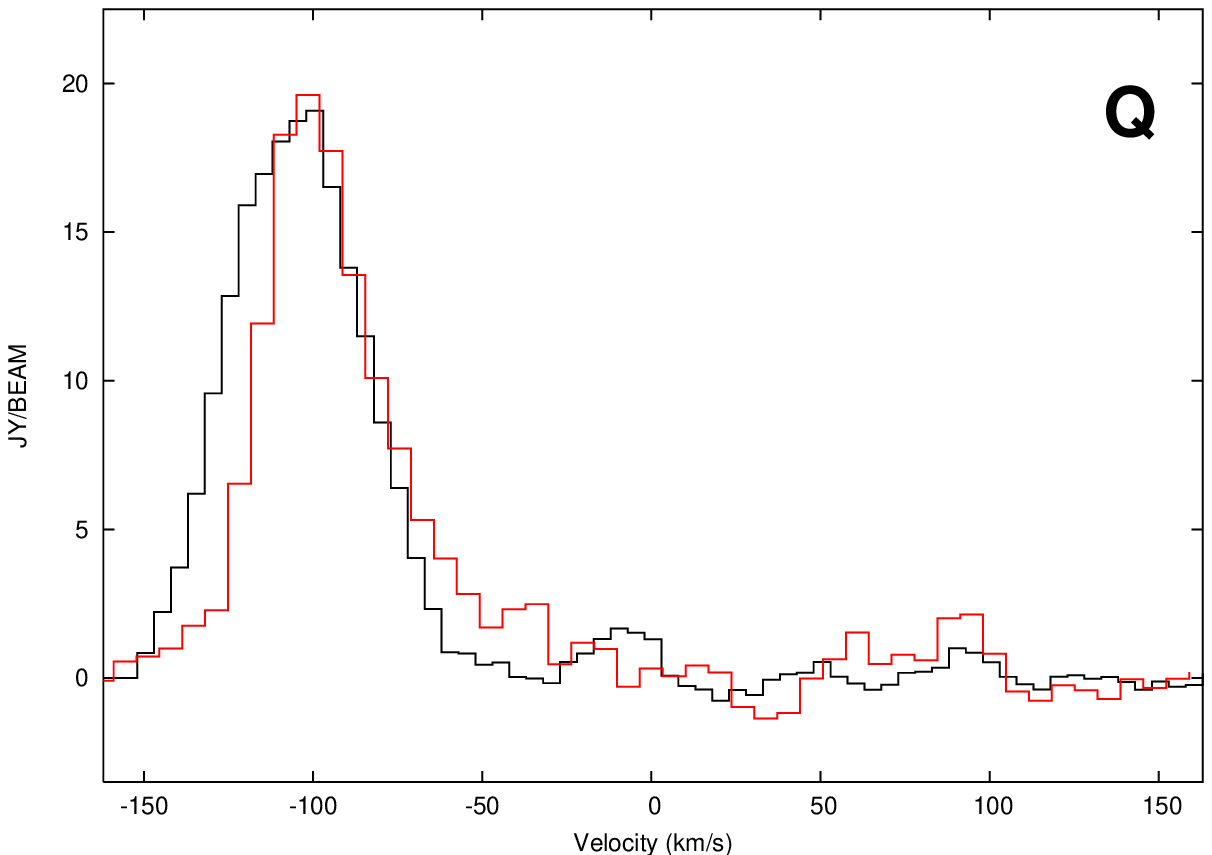}

\plotone{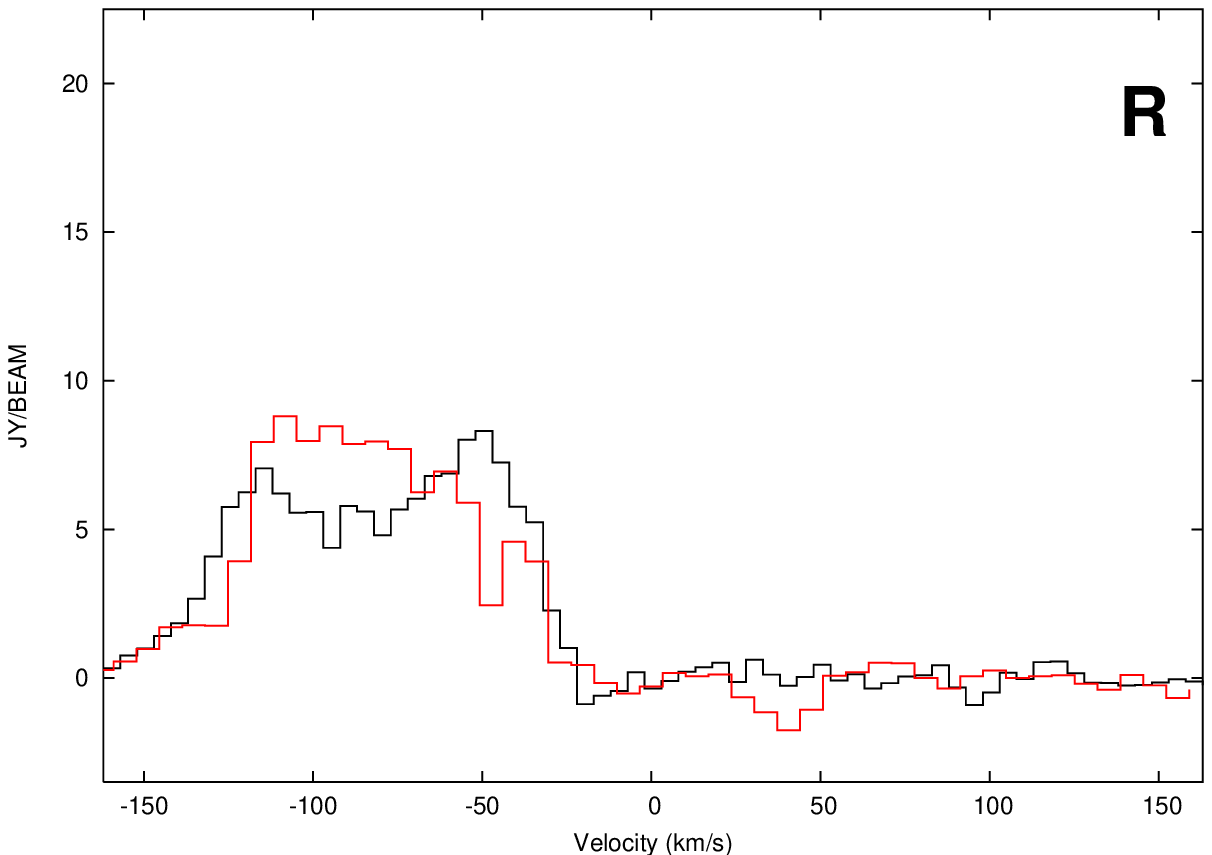}
\plotone{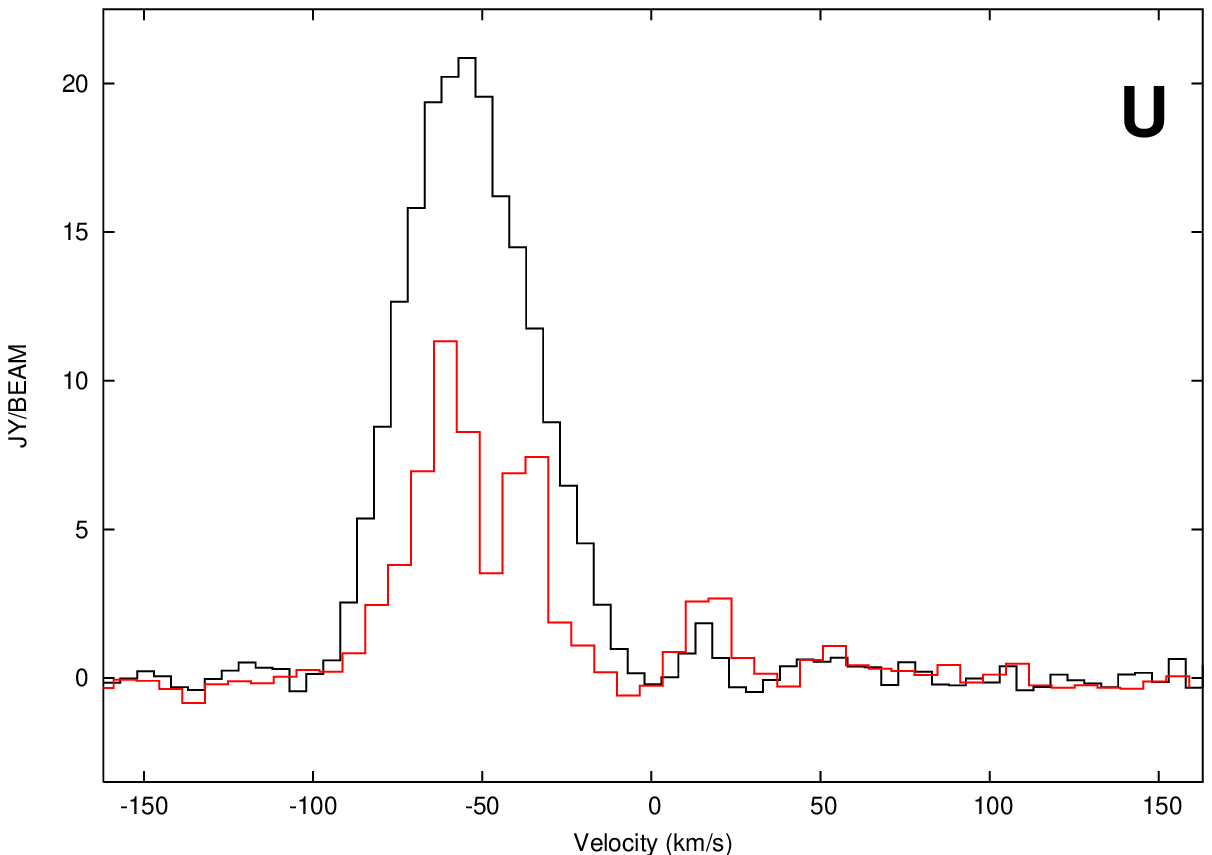}
\plotone{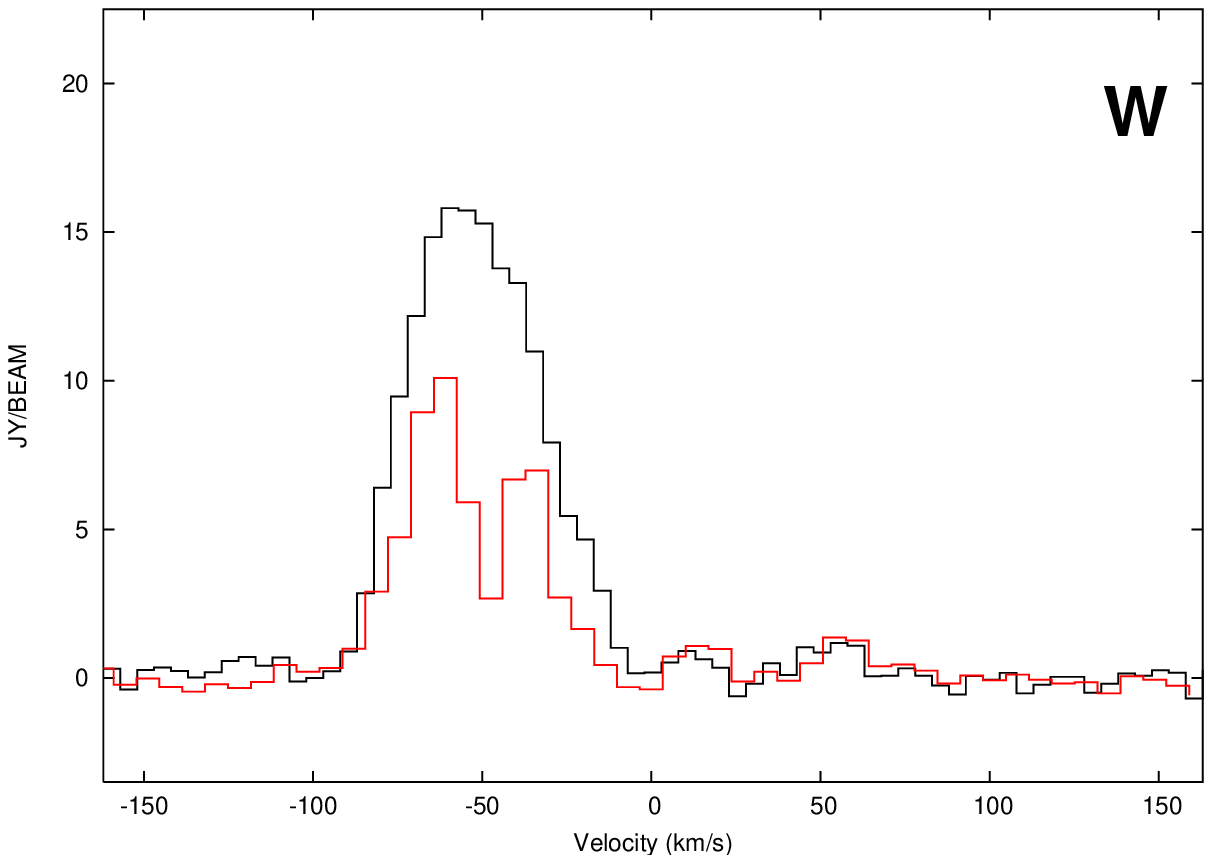}

\plotone{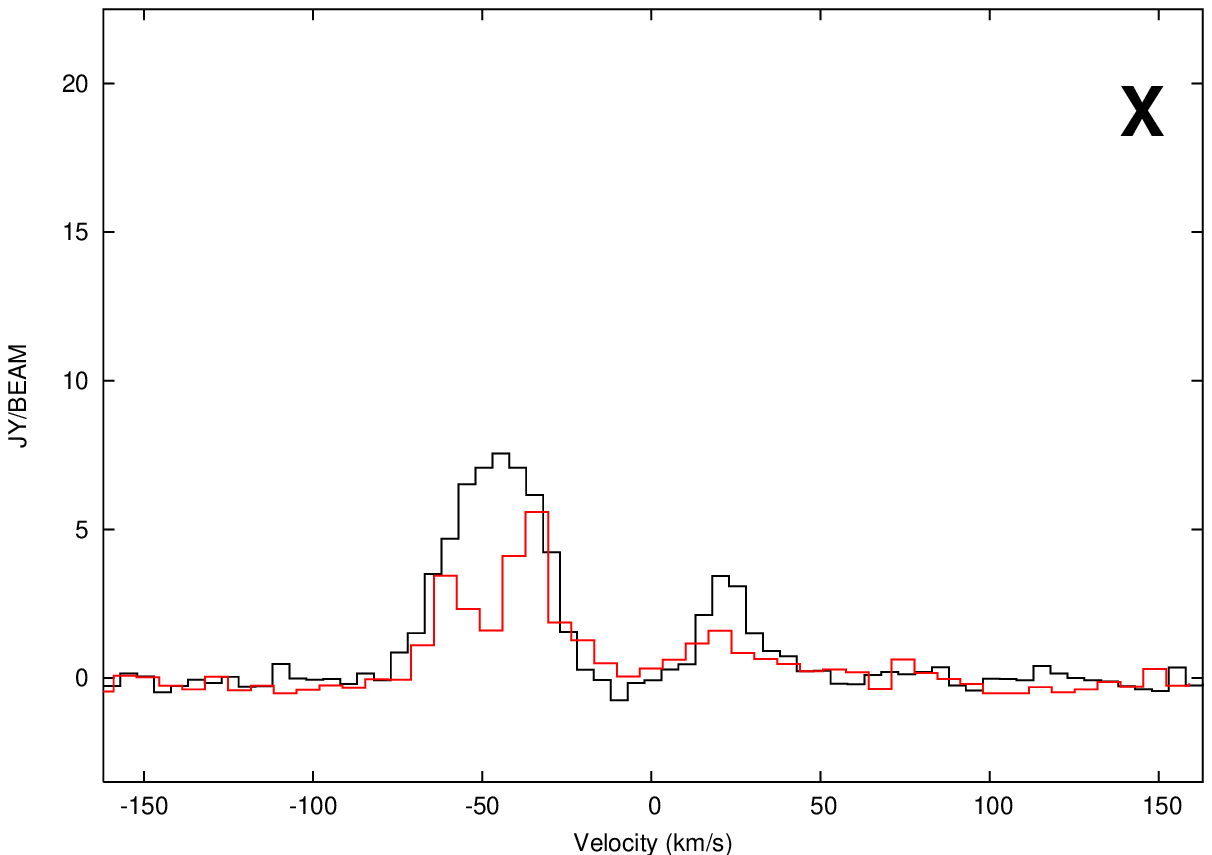}
\plotone{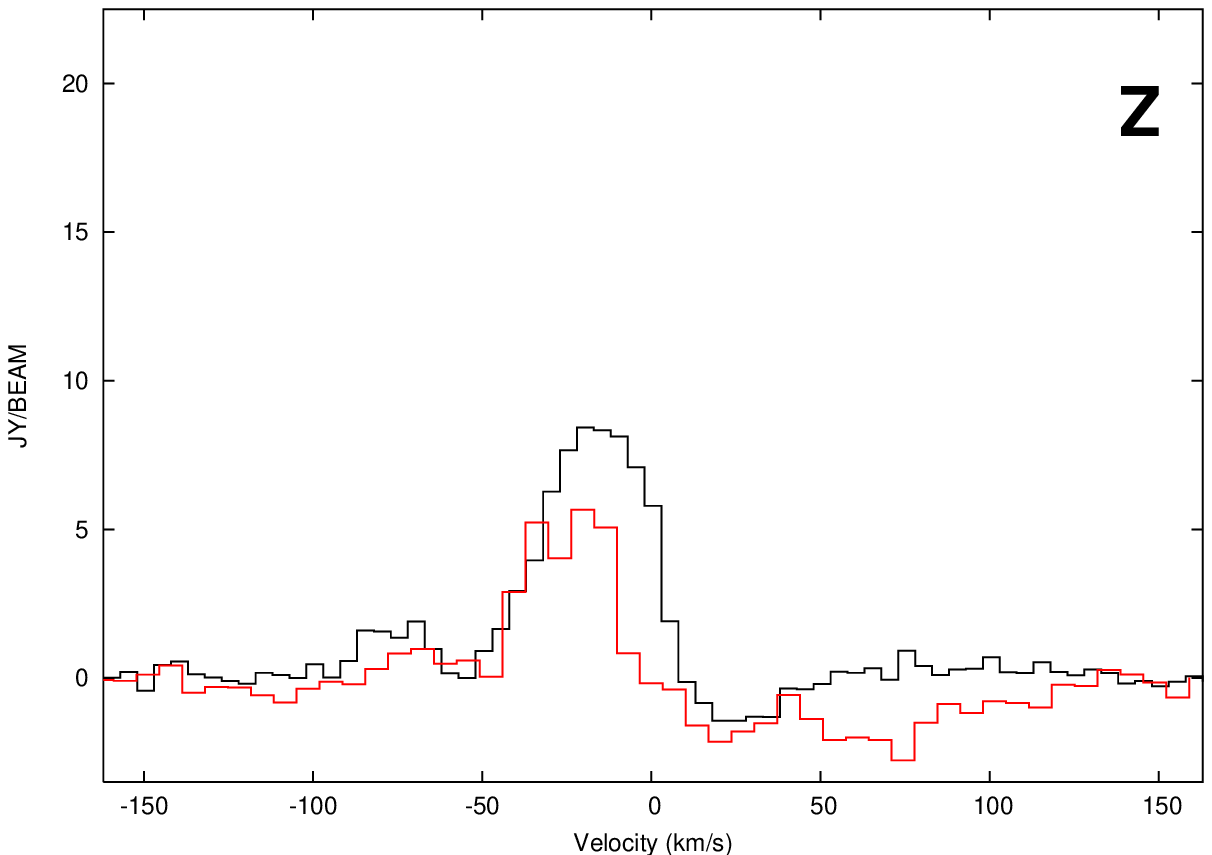}
\plotone{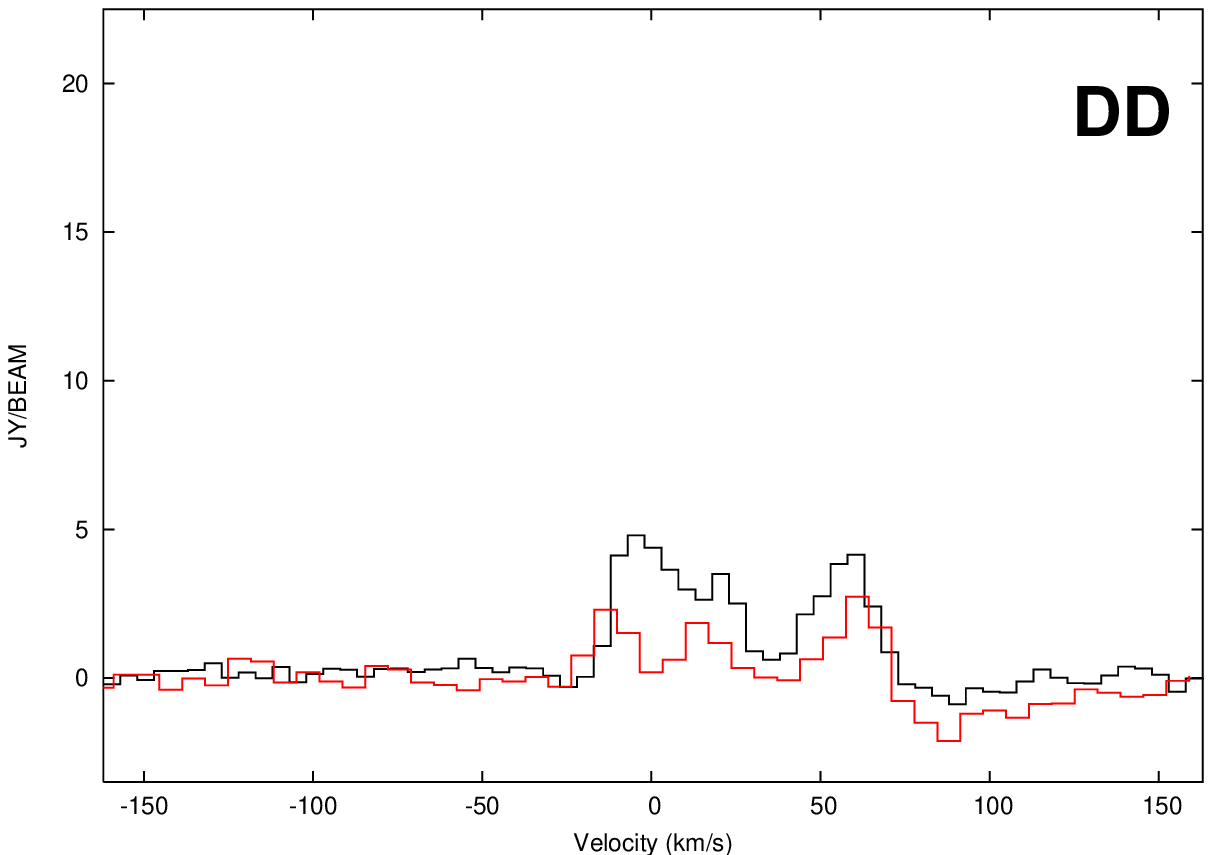}

\plotone{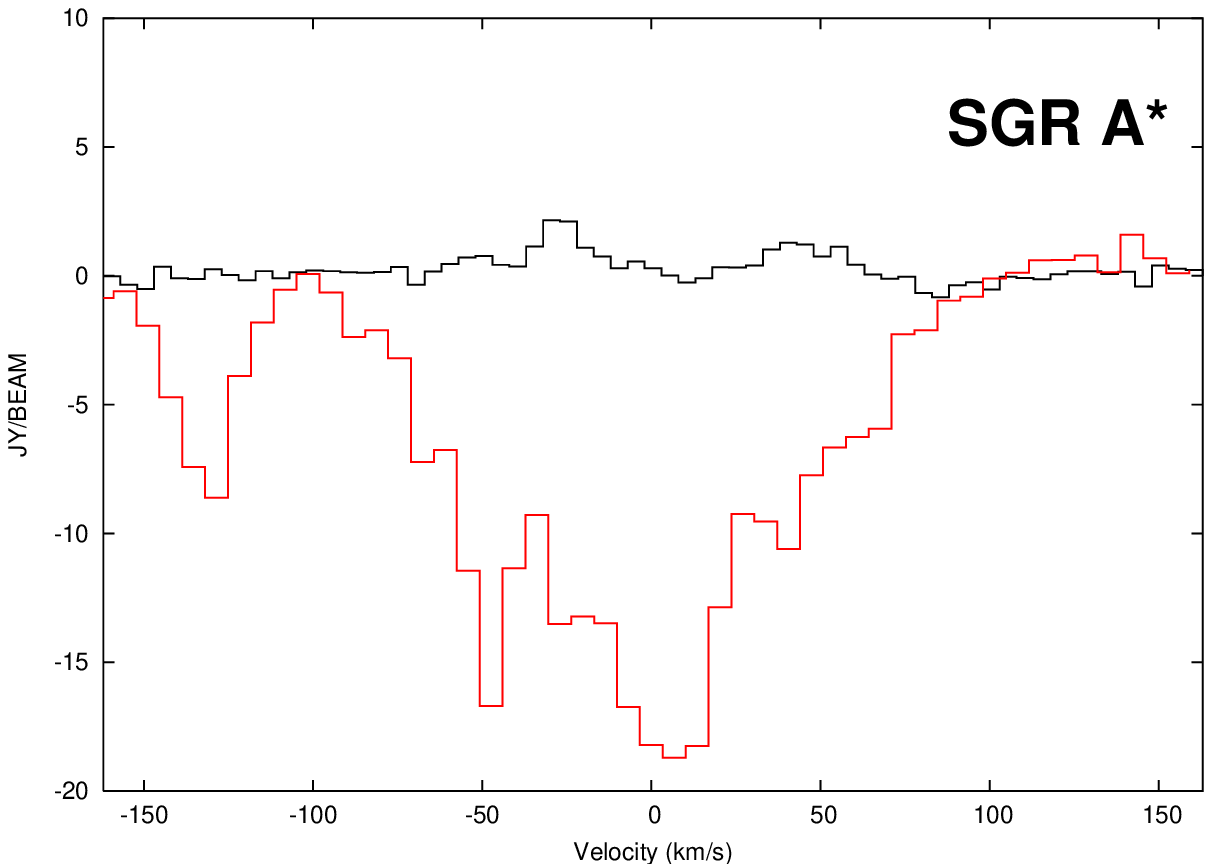}

\end{center}
\caption{HCN(4-3) spectra (black contours) overplotted to HCN(1-0) spectra (red contours) from \cite{chr05} at the locations of various clumps in figure \ref{spectra_hcn43.fig} and Sgr~A*(HCN(1-0) scaled up by a factor of 10 to fit on the plot). HCN(4-3) and HCN(1-0) convolved to the same resolution.\label{hcn10_spec.fig}}
\end{figure}

We overplot the spectra at the locations of various clumps along the CND for HCN(4-3) and HCN(1-0) from \cite{chr05} (figure \ref{hcn10_spec.fig}, convolved to the same resolution). We confirm that HCN(1-0) suffers from strong self-absorption, a problem especially noticeable in clumps K, N, U, X and W. Clump N in both transitions seems affected, but HCN(1-0) suffers a much more dramatic ``dip'' (i.e. a sharp absorption feature within the emission profile) than HCN(4-3). The velocity at which the absorption affects the HCN(1-0) line ($\sim$~0~km~s$^{-1}$, produced by a well known cold cloud complex, the ``local gas''; \citealt{gus87}) is slightly different from the velocity at which we find the dip in the HCN(4-3) spectrum. Also, the nearby clump K shows a similarly non-gaussian profile, but the rest of the clumps are mostly unaffected. We also overplot the spectra at the location of Sgr~A* (figure \ref{hcn10_spec.fig}, where we can observe the absorption features in the HCN(1-0) data noted by \cite{chr05}, which are absent from the HCN(4-3) data. Therefore, HCN(4-3) is probably not affected by absorption (and, if it is, not enough to explain the line profiles), but by the lack of short-spacings, a problem that can be solved by combining the interferometric data with single-dish data. HCN(1-0) could be missing extended emission as well, since the OVRO data have not been combined with single-dish data. Therefore, both transitions are probably affected by the short-spacings problem, but only in the case of (1-0) the self-absorption seems worrisome.

\begin{figure}
\begin{center}
\epsscale{1}
\plotone{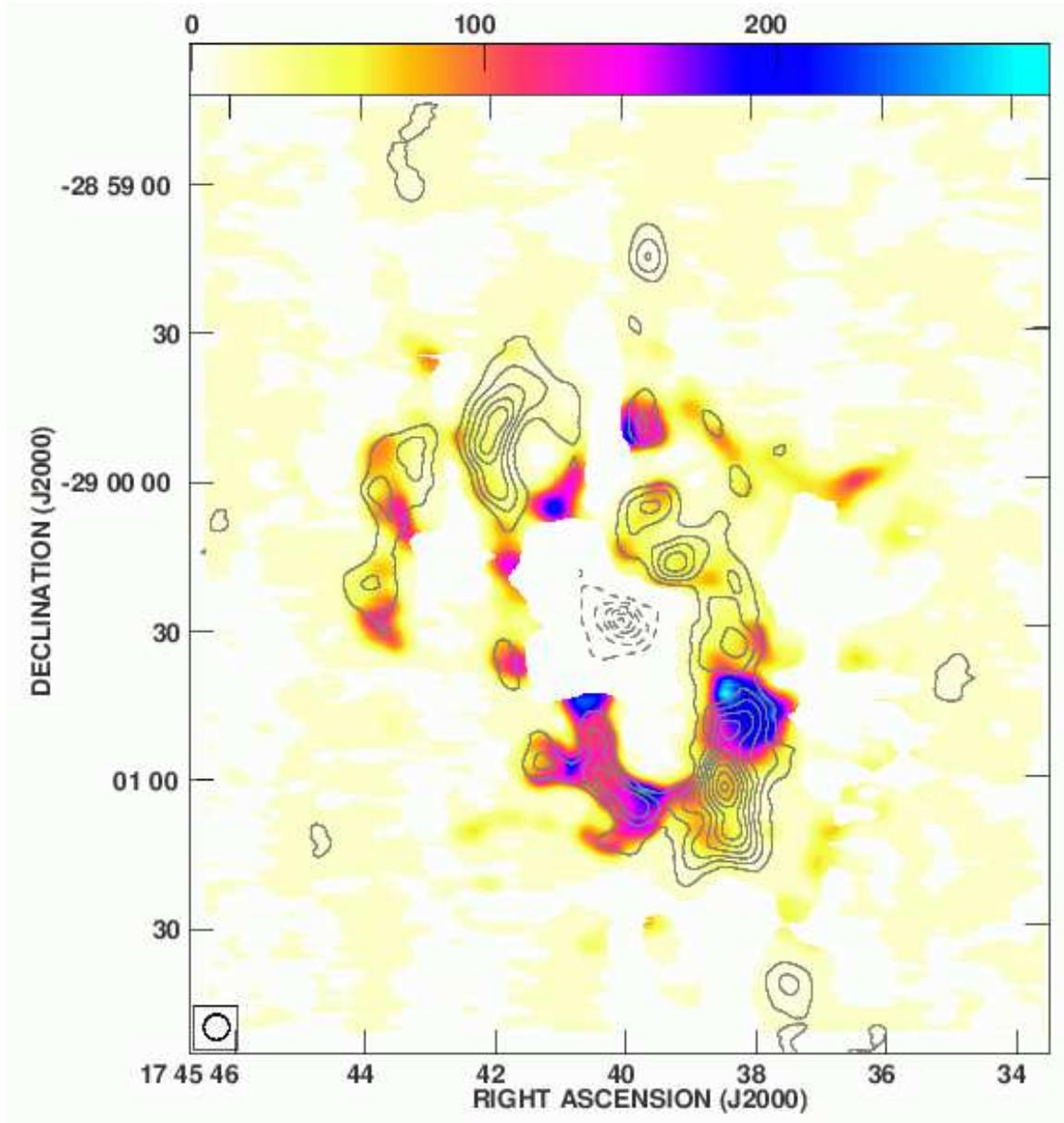}  
\end{center}
\caption{Ratio of HCN(4-3) and HCN(1-0) integrated intensity in color-scale. HCN(1-0) integrated intensity in contours. HCN(1-0) data from \cite{chr05}. Emission contour levels (solid lines) are in steps of 3$\sigma$, from 3$\sigma$ to 24$\sigma$, except for the highest contour level, at 26$\sigma$ (1.5~Jy~beam$^{-1}$~km~s$^{-1}$ to 13~Jy~beam$^{-1}$~km~s$^{-1}$). Absorption contour levels (dashed lines) are in steps of 10$\sigma$, from -10$\sigma$ to -60$\sigma$ (-5~Jy~beam$^{-1}$~km~s$^{-1}$ to -30~Jy~beam$^{-1}$~km~s$^{-1}$).\label{ratio43_10.fig}}
\end{figure}

The ratio of both transitions (figure \ref{ratio43_10.fig}, where the HCN(1-0) contours have been overplotted to better explain the ratio) confirms that the southern part of the CND is more highly excited than the northern part. We have calculated the ratio using only the points with flux greater than or equal to 3$\sigma$, and in the case where the flux was lower, we have taken 3$\sigma$ as the minimum value. The reasoning behind this treatment of the data was to assess the properties of the northern part of the CND. From figure \ref{hcn4_3_hcn.fig} we can observe that the emission from HCN(4-3) in the northern lobe is fairly extended. Since no emission from HCN(1-0) was observed in that same location, the ratio cannot be calculated without assuming an upper limit. Comparing the more excited molecular gas closer to the center with the less excited gas in the northern part of the CND would not have been possible. We have, however, not applied the same rule to the absorption features, leaving them blank, thus the void at the position of Sgr~A*. The missing flux densities problem does introduce some bias in the ratio, however, it is tipically not too severe and it is only for local spots. The overplotting of the HCN(1-0) integrated emission map in figure \ref{ratio43_10.fig} demonstrates that except for the {\it southern extension}, most of the regions with a high ratio correspond to the peripheral regions where the HCN(1-0) flux value has been artificially increased. This means for the hot regions in the northeast side of the CND, where the HCN emission is getting fainter, the emission must be hotter than what we are able to show in the ratio map. Where both HCN(1-0) and HCN(4-3) have been detected, for example the {\it northeast lobe}, the value of the ratio is one of the lowest values along the CND. The ratio map therefore shows the {\it southern extension} as a whole, to be more excited that any other part of the CND. However, the absence of detection in the eastern part of the CND cannot rule out the presence of hotter gas in the region, which could possibly be more diffuse. The large ratios in other parts of the CND, as in the {\it southwest lobe}, are interspersed among regions with smaller ratios. This is consistent with heating at the boundaries of clumps. Although all of the southern part of the CND is really affected by self-absorption (figure \ref{hcn10_spec.fig}), this is an interesting result since it suggests that the {\it southern extension} is formed in its entirety by more excited gas than other parts of the CND. The question arises, then, why is the southeastern part of the CND more excited? What is the heating mechanism? Shocks? Radiation? We will address these questions later, with the help of an extra comparison tool, a high-temperature molecular tracer, NH$_3$. At the same time, the ratio map, especially in the northern part of the CND, suggests that the ratio increases as the gas approaches the center, i.e. the inner part of the CND is more excited than the outer part. \cite{har85} suggested that the UV radiation from the nuclear cluster (which is located in a cavity void of dust, and therefore transparent to the UV radiation) could be responsible for the heating of the molecular gas composing the CND. The inner edge of the molecular ring would be heated by this radiation, but as the distance from the center increases, the radiation starts to be absorbed by the dust also composing the ring. Thus, the outer parts of the CND would be less affected by this radiation. Finally, it is interesting to remark that the location of clump E, where we found a broad double-peak profile (figures \ref{spectra_hcn43.fig} and \ref{hcn10_spec.fig}) shows a large ratio value. This result suggests that clump E has a higher excitation than the surrounding environment. The interaction with the northern arm of the minispiral may be a possible mechanism. A detailed kinematic study of the velocity distribution in the minispiral is needed to fully elucidate this situation, possibly.

When comparing the HCN(4-3) emission map with emission maps produced by NH$_3$ ((3,3) and (6,6), \cite{mcg01} and \cite{her02}, respectively) we can gain a better understanding of the spatial distribution of the high-density molecular gas in the inner 2~pc of the Galactic center. NH$_3$ has a lower critical density (10$^{5}$~cm$^{-3}$) as compared to HCN, but traces gas at higher temperatures. Both NH$_3$(3,3) and (6,6) have remarkably different distributions in the central region of the Milky Way as shown by \cite{mcg01} and \cite{her02,her05}.

\begin{figure}
\begin{center}
\epsscale{1}
\plotone{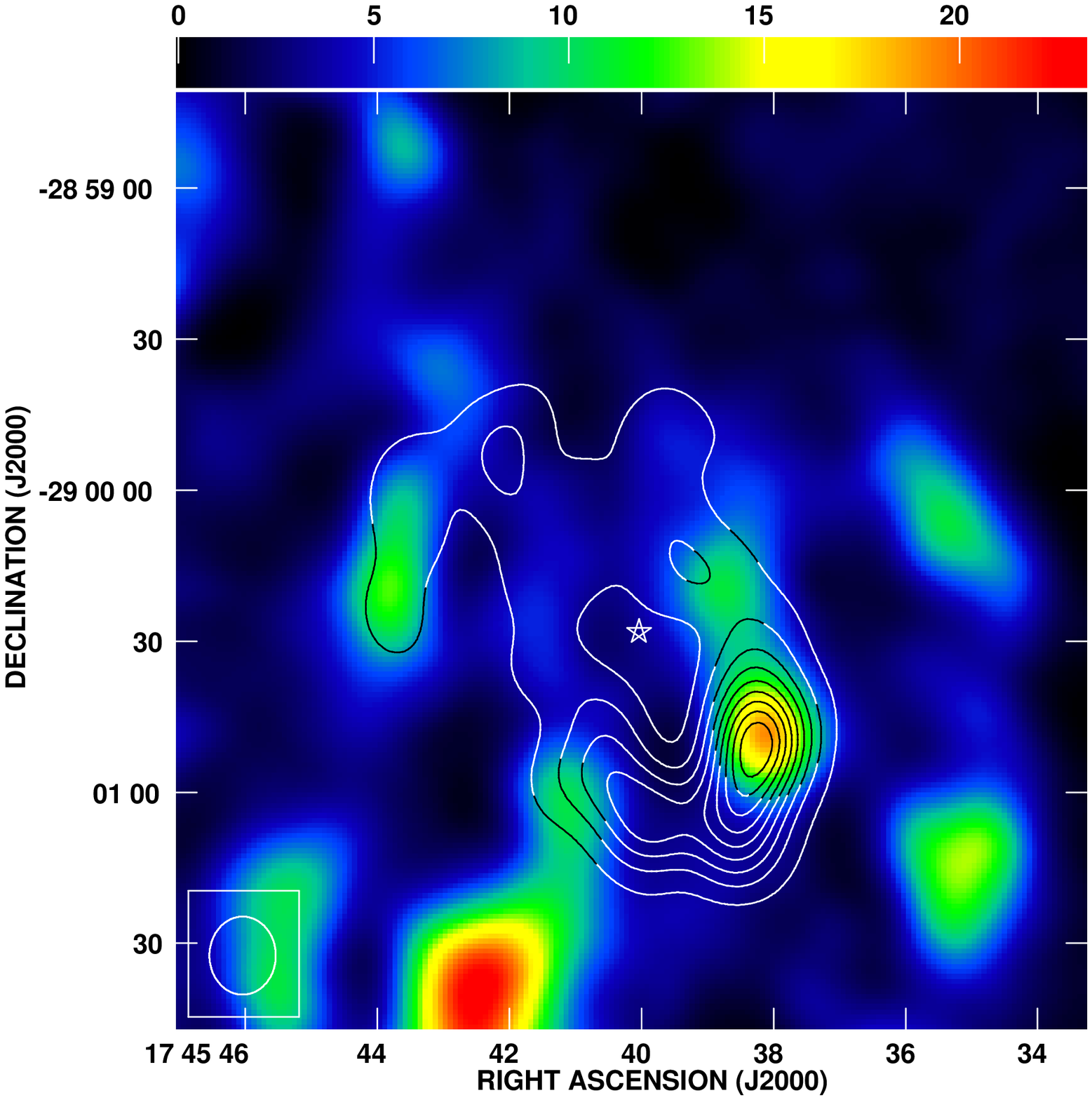}
\end{center}
\caption{HCN(4-3) integrated intensity in contours. NH$_3$(3,3) integrated intensity in false color-scale from \cite{mcg01}. The HCN(4-3) image has been smoothed to match the resolution of the NH$_3$ image. Contour levels are are in steps of 10\% of the intensity peak, from 1.8~$\times$~10$^{2}$ to 16.2~$\times$~10$^{2}$~Jy~beam$^{-1}$~km~s$^{-1}$. The false color-scale is in ~Jy~beam$^{-1}$~km~s$^{-1}$. Sgr~A* is marked with a star.\label{hcn_33.fig}}
\end{figure}

A comparison of HCN(4-3) to NH$_3$(3,3) (figure \ref{hcn_33.fig}; HCN(4-3) smoothed to match the NH$_3$(3,3) resolution) shows that the emission distributions from these two molecular lines are poorly correlated. However, both lines trace the western side of the CND and show the strongest peak along the CND at the same position on the {\it southwest lobe}. HCN(4-3) and NH$_3$(3,3) emission coincide in the {\it eastern arm} as well. It is worth noting that the {\it southern streamer} as traced by NH$_3$(3,3) (at $\alpha_{J2000.0}$~=~17$^h$45$^m$43$^s$ and $\delta_{J2000.0}$~=~-29$\arcdeg$01$\arcmin$40$\arcsec$) seems to connect to the eastern side of the CND as traced by HCN(4-3). \cite{her05} reported that no kinematic connection could be made between the {\it southern streamer} and the CND, but at the same time, it was clear that there was a change in the NH$_3$(3,3) emission when (at least in projection) it reached the CND. Therefore, the spatial alignment at the CND might only be a projection effect along the line of sight, but a physical connection could not be completely ruled out because of the resolution of the NH$_3$(3,3) data. We do indeed observe the same trend. The gas traced by both NH$_3$(3,3) and HCN(4-3) in the eastern part do not overlap, but it seems that the emission from NH$_3$(3,3) does become weaker in the vicinity of the CND. At present sensitivity, the kinematic and structural information still cannot elucidate whether there is a connection between the {\it southern streamer} and the CND.

\begin{figure}
\begin{center}
\epsscale{1}
\plotone{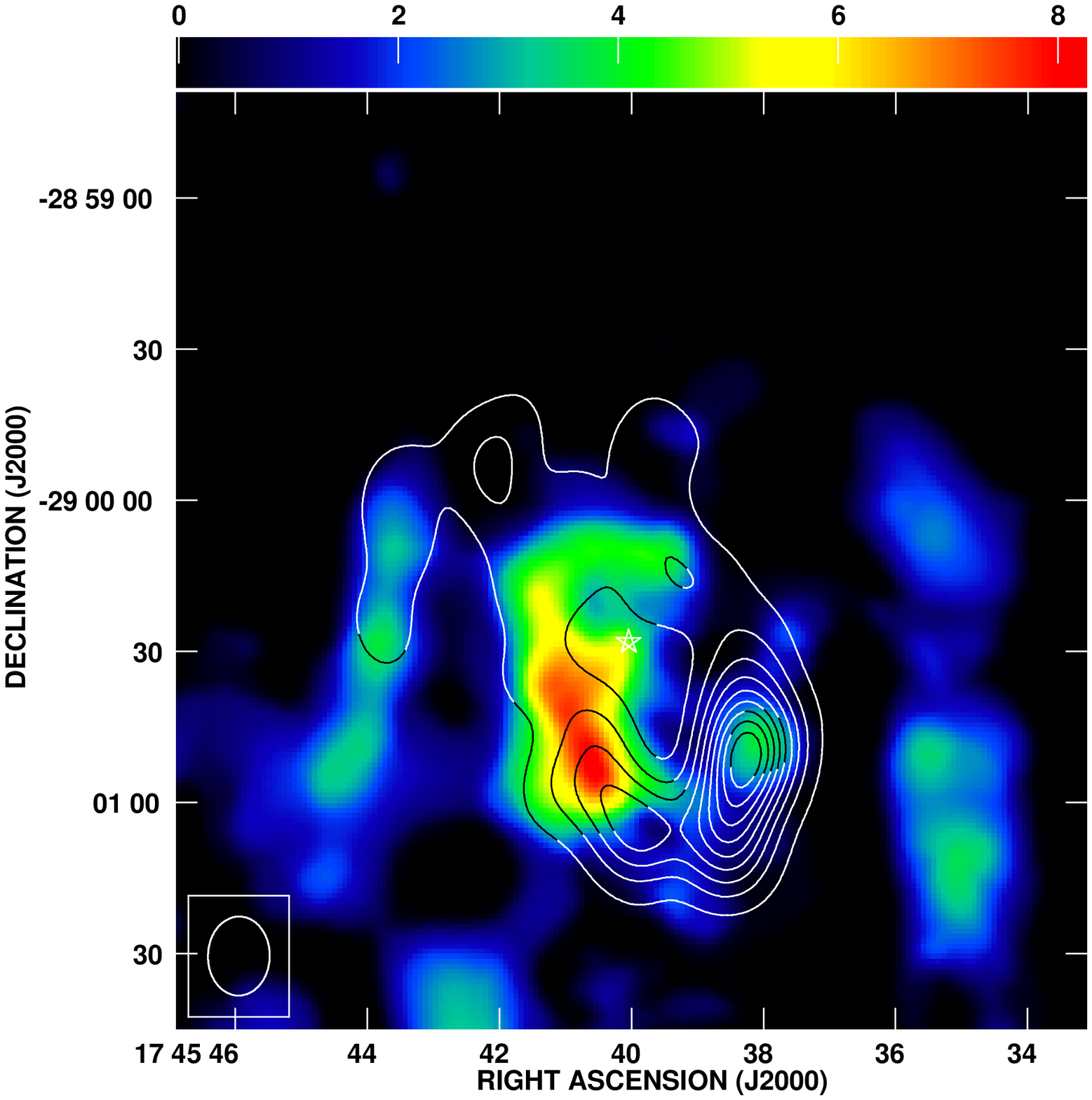}
\end{center}
\caption{HCN(4-3) integrated intensity in contours. NH$_3$(6,6) integrated intensity in false color-scale from \cite{her02}. The HCN(4-3) image has been smoothed to match the resolution of the NH$_3$ image. Contour levels are in steps of 10\% of the intensity peak, from 1.8~$\times$~10$^{2}$ to 16.2~$\times$~10$^{2}$~Jy~beam$^{-1}$~km~s$^{-1}$. The false color-scale is in ~Jy~beam$^{-1}$~km~s$^{-1}$. Sgr~A* is marked with a star.\label{hcn_66.fig}}
\end{figure}

The comparison of HCN(4-3) and NH$_3$(6,6) (figure \ref{hcn_66.fig}) shows a good correspondence in the eastern and northern parts of the CND, and even in the {\it northeast arm}. However, we can see that NH$_3$(6,6) is detected inside the central cavity of the CND, apparently approaching Sgr~A*, while HCN(4-3) is only detected along the CND. This result suggests that the molecular gas that composes the CND is in general colder and perhaps denser than the gas that is located inside of the cavity.

\begin{figure}
\begin{center}
\epsscale{1}
\plotone{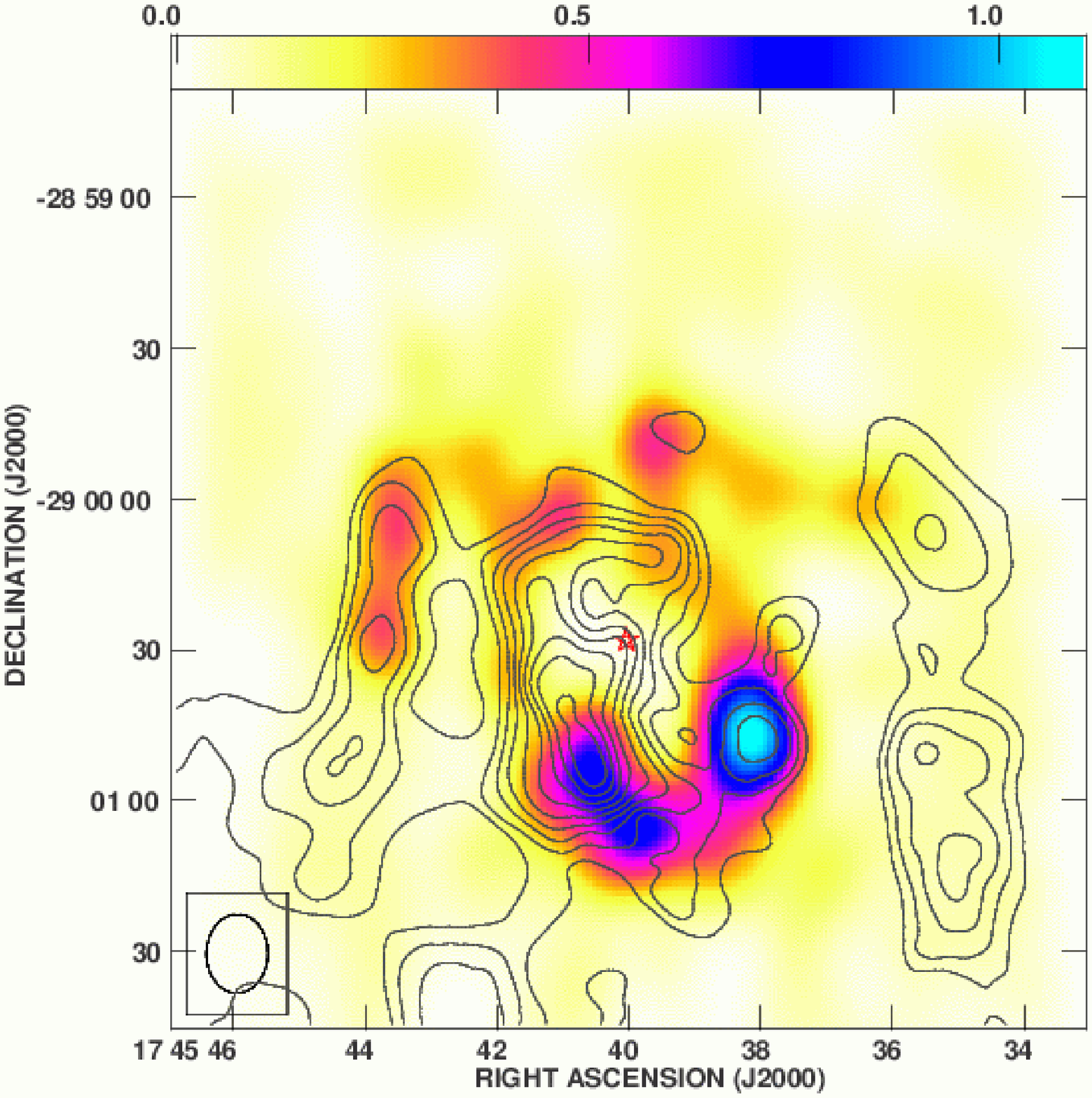}
\end{center}
\caption{Ratio of HCN(4-3) and (1-0) in color-scale. NH$_3$(6,6) integrated intensity in contours from \cite{her02}. Contour levels are in steps of 10\% of the intensity peak, from 8~$\times$~10$^{-1}$ to 74~$\times$~10$^{-1}$~Jy~beam$^{-1}$~km~s$^{-1}$. Sgr~A* is marked with a star.\label{66_ratio.fig}}
\end{figure}

By overplotting the HCN(4-3)/HCN(1-0) ratio and the NH$_3$(6,6) integrated emission we can make an interesting comparison (figure \ref{66_ratio.fig}). We observe that the southeastern part of the CND is located south of the strongest peak detected in NH$_3$(6,6). It looks like the material traced by NH$_3$(6,6) is ``following'' the path marked by the ratio towards Sgr~A*. Since the large value of the ratio in the southeastern part of the CND indicates the presence of a larger amount of highly-excited material than of low-excited material, it means that the majority of the molecular gas in that region is warmer. Because NH$_3$(6,6) traces very warm gas and it penetrates from the eastern side of the CND towards the supermassive black hole, the southeastern side of the CND could be becoming warmer as it flows northwest, towards Sgr~A*.

\begin{figure}
\begin{center}
\epsscale{1}
\plotone{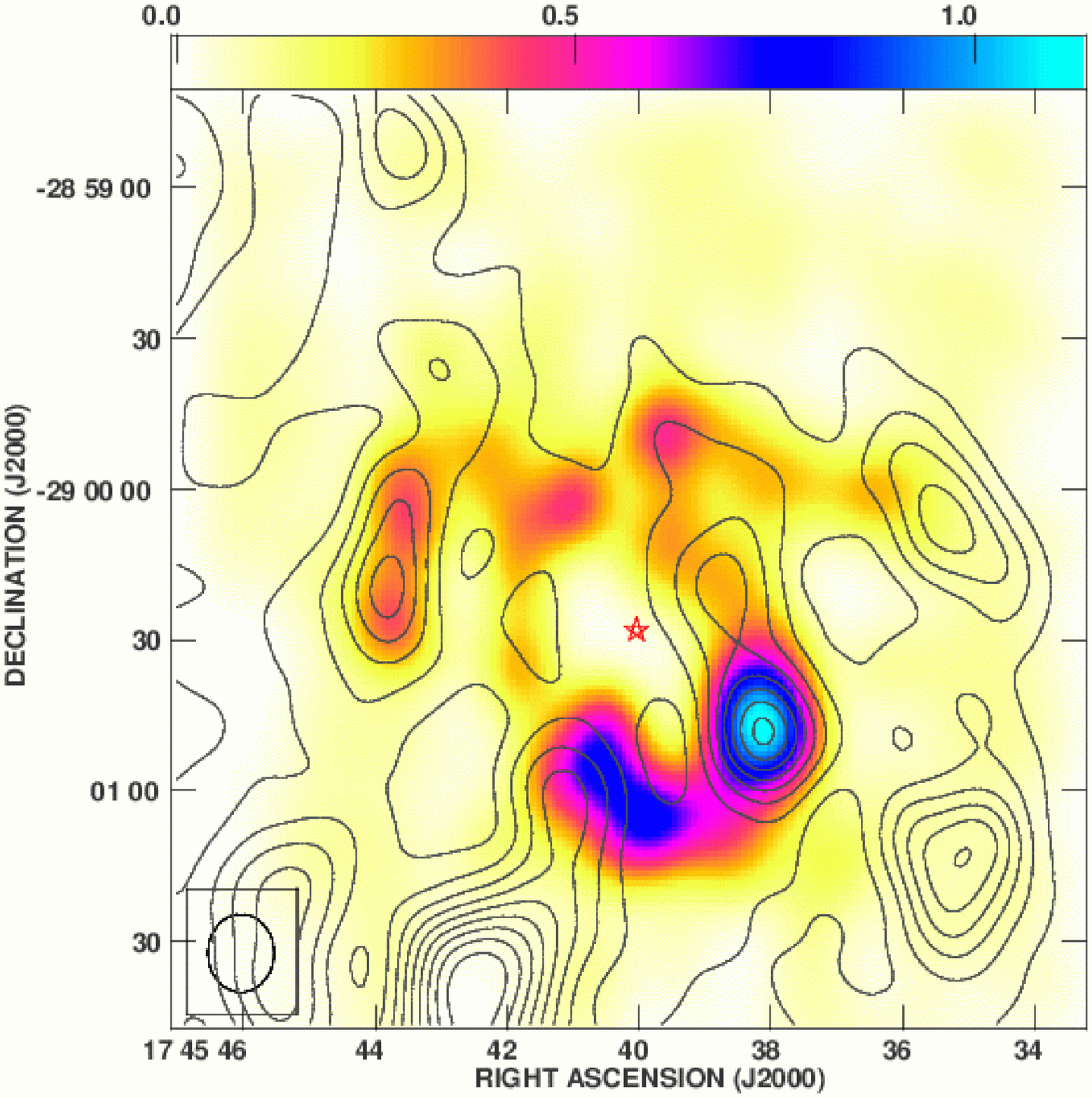}
\end{center}
\caption{Ratio of HCN(4-3) and (1-0) in color-scale. NH$_3$(3,3) integrated intensity in contours from \cite{mcg01}. Contour levels are in steps of 10\% of the intensity peak, from 8~$\times$~10$^{-1}$ to 74~$\times$~10$^{-1}$~Jy~beam$^{-1}$~km~s$^{-1}$. Sgr~A* is marked with a star.\label{33_ratio.fig}}
\end{figure}

At the same location where the gas traced by NH$_3$(6,6) heads for Sgr~A*, the value of the ratio changes radically, as if suddenly the level of excitation in the gas tracing the CND drops. At that location HCN(1-0) could be slightly affected by self-absorption (spectrum Z in figure \ref{hcn10_spec.fig}), but the very low value of the ratio indicates that the amount of HCN(4-3) is also low, otherwise the ratio would have been large. The question remains as to why the material is more highly excited precisely in the {\it southern extension}. When plotting the HCN(4-3)/HCN(1-0) ratio with the NH$_3$(3,3) integrated intensity map from \cite{mcg01} (figure \ref{33_ratio.fig}), we observe that the region where the {\it southern streamer} seems to reach the CND coincides with the location of the northernmost part of the high ratio, where NH$_3$(6,6) becomes stronger before heading northwest. Consequently, in terms of projection, there is a location in the southeastern part of the CND where the {\it southern streamer} traced by NH$_3$(3,3), the strongest peak in the NH$_3$(6,6) emission and the edge of the high-ratio area all coincide. The {\it southern streamer} may impact upon the CND producing destabilized material which infalls towards the supermassive black hole. This material may become so highly excited that HCN(4-3) energy levels may be depopulated so that a higher HCN transition will be needed to trace the infalling gas.

\begin{figure}
\begin{center}
\epsscale{1}
\plotone{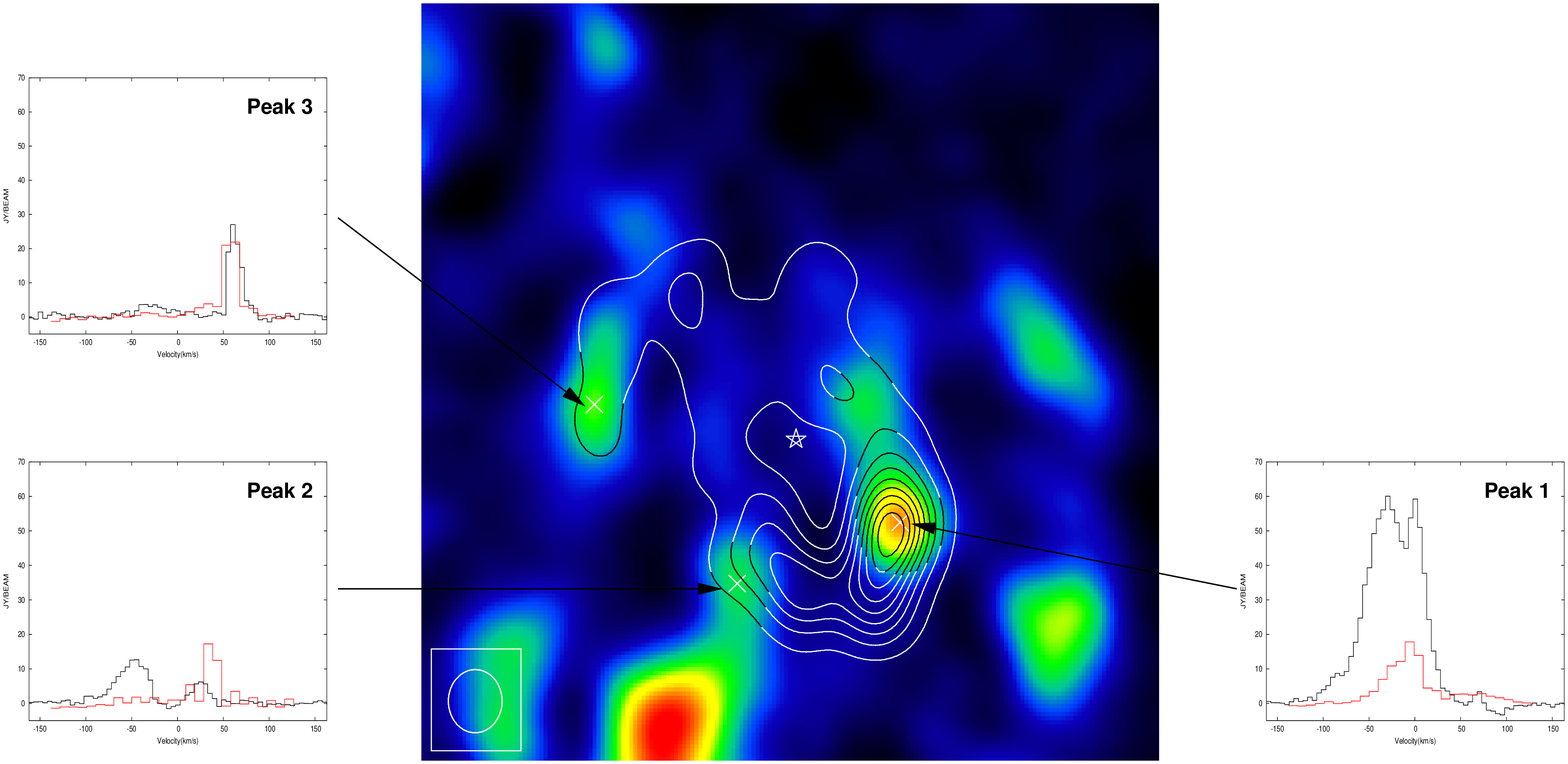}
\end{center}
\caption{HCN(4-3) spectra (black contours) overplotted to NH$_3$(3,3) spectra (red contours) from \cite{mcg01} at the locations of various peaks. The NH$_3$(3,3) data have been scaled up by a factor of 50 to fit on the plot. The false color-scale is in ~Jy~beam$^{-1}$~km~s$^{-1}$. Sgr~A* is marked with a star.
\label{33_spec.fig}}
\end{figure}

\begin{figure}
\begin{center}
\epsscale{1}
\plotone{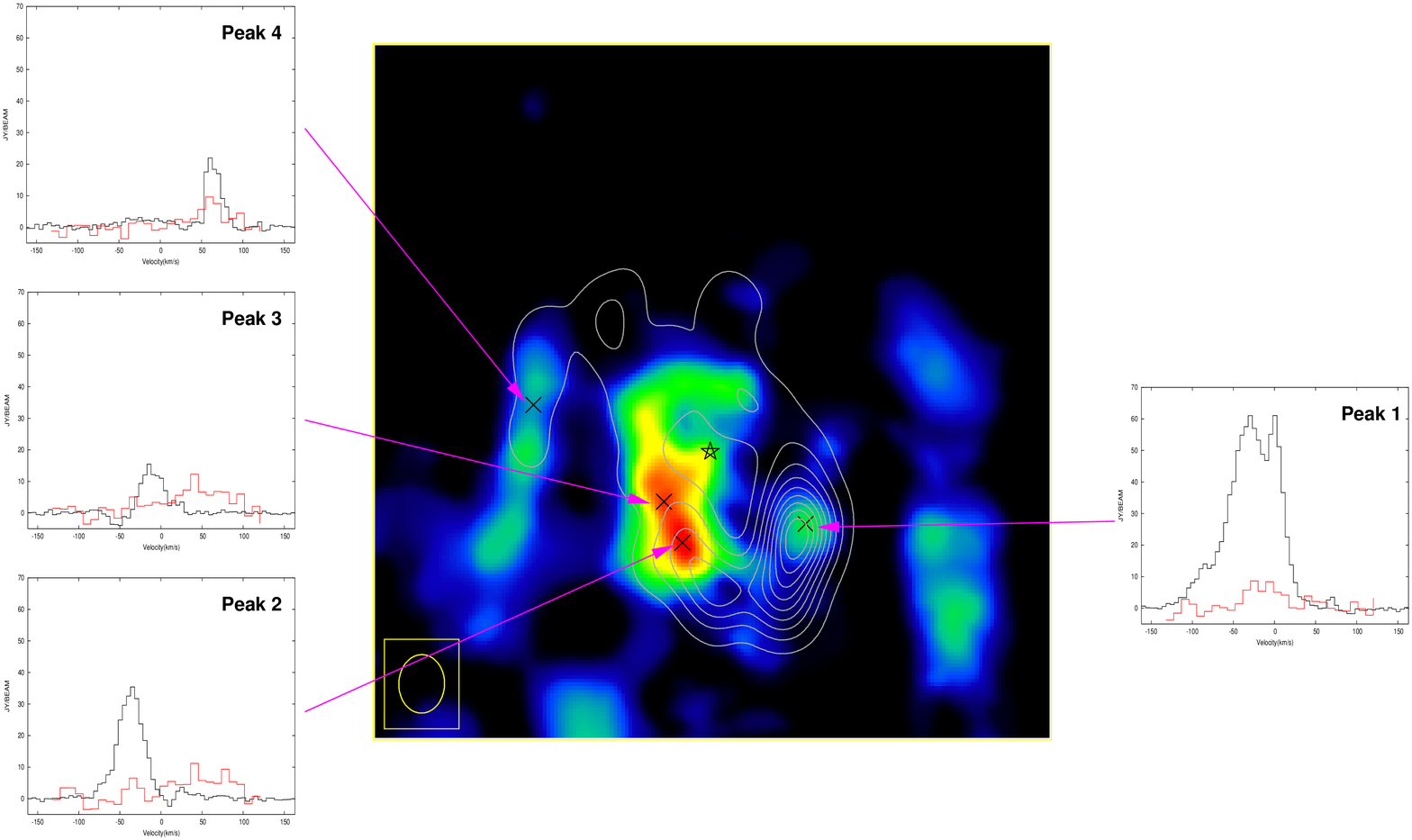}
\end{center}
\caption{HCN(4-3) spectra (black contours) overplotted to NH$_3$(6,6) spectra (red contours) from \cite{her02} at the locations of various peaks. The NH$_3$(6,6) data have been scaled up by a factor of 100 to fit on the plot. The false color-scale is in ~Jy~beam$^{-1}$~km~s$^{-1}$. Sgr~A* is marked with a star.
\label{66_spec.fig}}
\end{figure}

A comparison of spectra of both the HCN(4-3) and the NH$_3$ data is needed to determine whether the material traced by these transitions is the same. In order to accomplish this study we take the spectra at various locations (figures \ref{33_spec.fig} and \ref{66_spec.fig}). From figure \ref{33_spec.fig} we find that the material tracing the {\it southwest lobe} and the {\it northeast arm} (peaks 1 and 3) kinematically coincides both in HCN(4-3) and NH$_3$(3,3), but the same cannot be said about peak 2, located on the region where the material from the ``20 km~s$^{-1}$ GMC'' approaches the CND. However, we do detect a small peak around 20~km~s$^{-1}$, weaker than the peak detected around -50~km~s$^{-1}$, but nonetheless present, which indicates the detection of material from the {\it southern streamer}. We observe a similar situation in figure \ref{66_spec.fig} when comparing HCN(4-3) and NH$_3$(6,6). The same material is detected by HCN(4-3) and NH$_3$(6,6) in the {\it southwest lobe} and the {\it northeast arm} (peaks 1 and 4) but not in the regions where the material approaches Sgr A* (peaks 2 and 3). Therefore, the material detected in the northern and western parts of the CND by HCN(4-3) and NH$_3$(3,3) and (6,6) coincides, indicating that the denser and colder gas is heavily mixed with the more diffuse and warmer gas at those locations. However, the material detected in the eastern part of the CND appears to be very different, with the gas approaching the black hole warmer and more diffuse than the gas tracing the CND and a possible interaction between the gas from the {\it southern streamer} and the gas forming the CND in the southeasternmost part of the ring. These results seem to support the previously noted conclusion, where the material in the {\it southern extension} is warmed and pushed towards the dynamical center of the Milky Way by the action of the gas coming from the ``20 km~s$^{-1}$ GMC''.

\section{CS(7-6)}

\begin{figure}
\begin{center}
\epsscale{1}
\plotone{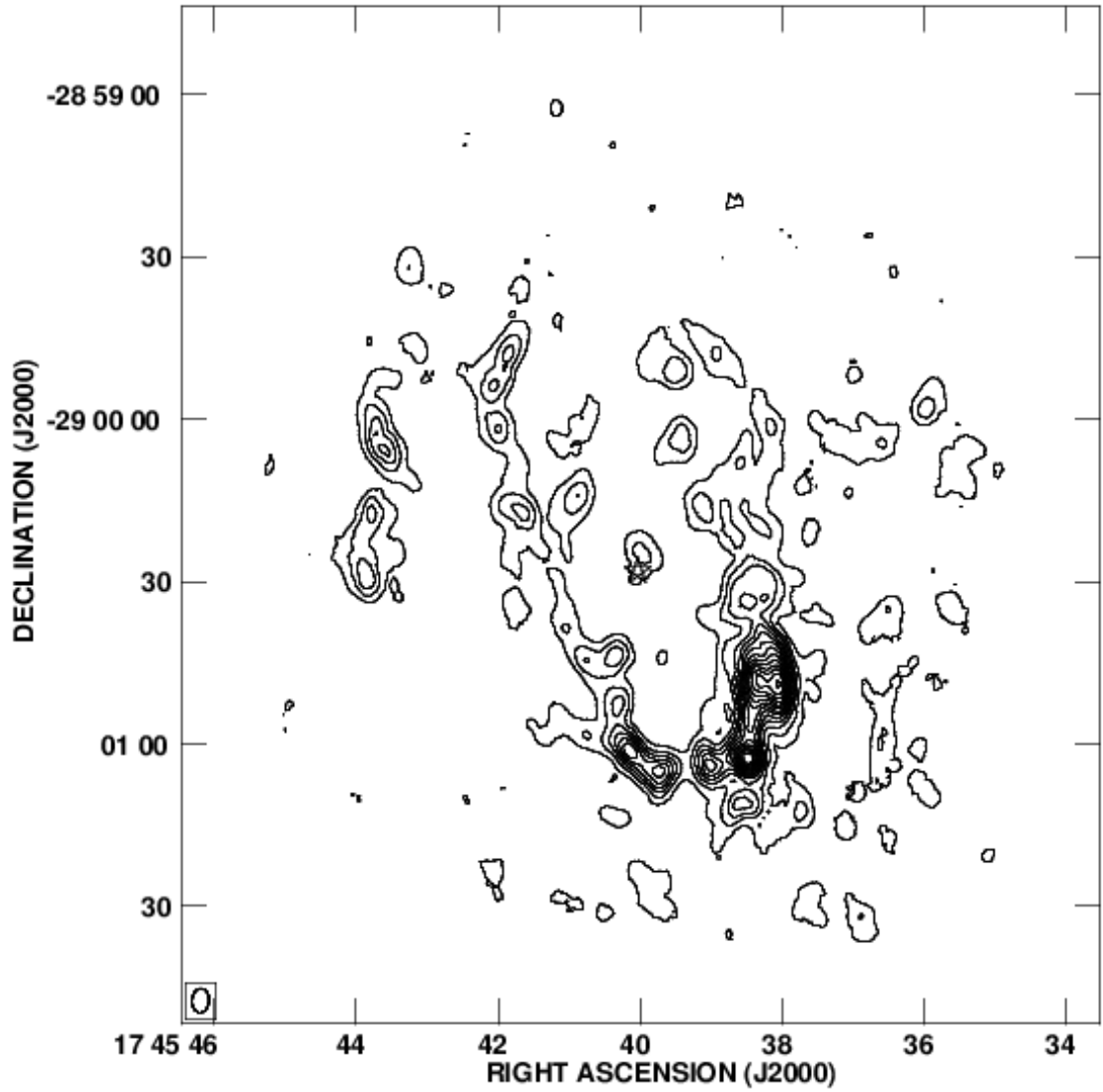}
\end{center}
\caption{CS(7-6) integrated intensity emission map. Contour levels are in steps of 6$\sigma$, from 3$\sigma$ to 69$\sigma$, except for the highest contour level, at 76$\sigma$ (1.0$\times$~10$^{1}$ to 25.3$\times$~10$^{1}$~~Jy~beam$^{-1}$~km~s$^{-1}$). Sgr~A* is marked with a star.\label{cs7_6.fig}}
\end{figure}

We employed a correlator setup to sample the CS(7-6) line at the same time. Because of the locations of the two lines, CS(7-6) had to be placed at the edge of the passband and the velocity coverage was not as broad for this line, from -150 to 128~km~s$^{-1}$. As previously noted, this CS transition traces gas at even higher temperature than the observed HCN line but at a slightly lower density. We clearly detect and resolve the CS(7-6) emission in the southern part of the CND, but it is much weaker towards the northern part, although it is clearly detected in the {\it northeast arm} (figure \ref{cs7_6.fig}). This result is consistent with the HCN results in suggesting that the northern and the southern parts of the CND have different excitation levels, since the warmer gas is absent from the north.

\begin{figure}
\begin{center}
\epsscale{1}
\plotone{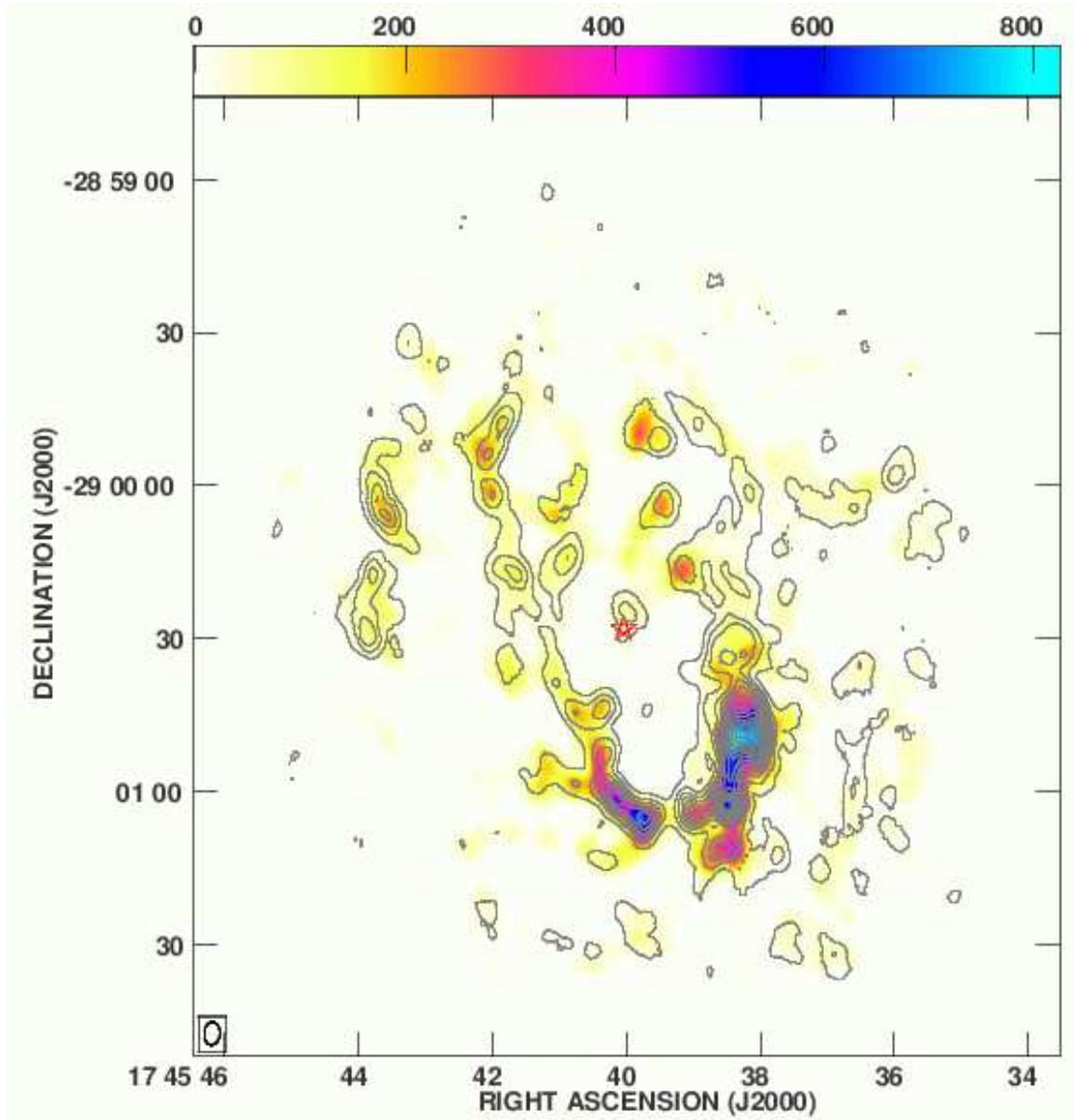}
\end{center}
\caption{CS(7-6) integrated intensity in contours. HCN(4-3) integrated intensity in false color-scale. Contour levels are as in figure \ref{cs7_6.fig}. The false color-scale is in ~Jy~beam$^{-1}$~km~s$^{-1}$. Sgr~A* is marked with a star.\label{cs7_6_hcn4_3.fig}}
\end{figure}

\begin{figure}
\begin{center}
\epsscale{1}
\plotone{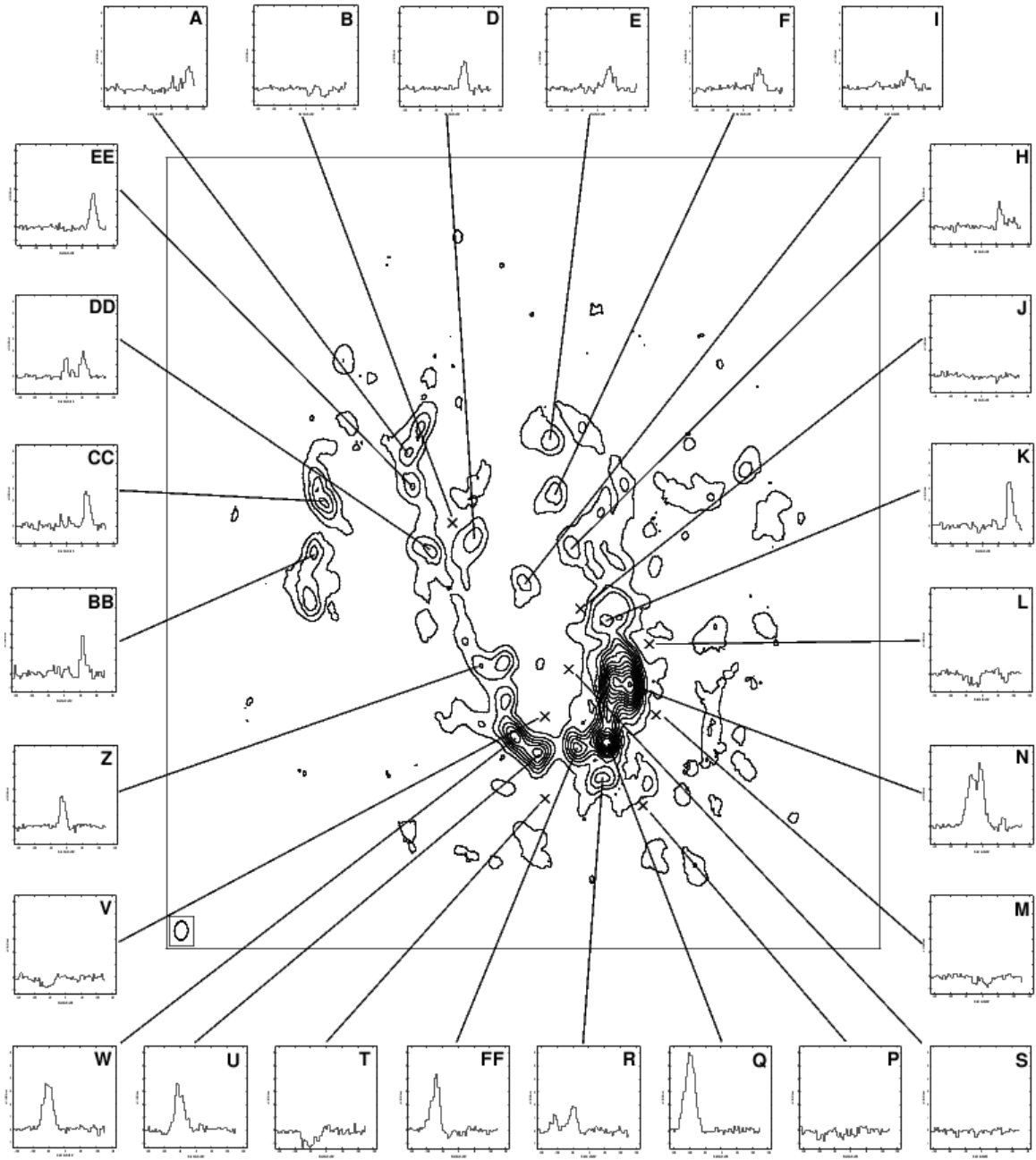}
\end{center}
\caption{CS(7-6) spectra measured at the positions of the different clumps. The clumps have been named as in figure \ref{spectra_hcn43.fig} (except clump FF) to facilitate the comparison between the two sets of spectra. Wherever the peak was not exactly at the same location the nearest peak has been used.\label{cs_spectra.fig}}
\end{figure}

The results for the CS(7-6) and HCN(4-3) lines show that these two molecualr lines correlate quite well with each other (figure \ref{cs7_6_hcn4_3.fig}). The emission peaks coincide in the southern part of the CND, and even in the few places where CS(7-6) is found in the northern part of the CND. Also, if we compare the spectra at the same positions (figures \ref{spectra_hcn43.fig} and \ref{cs_spectra.fig}, CS(7-6) spectra plotted in the same velocity range as HCN(4-3) spectra), we observe that both molecular lines have similar line profiles (FWHM~$\approx$~35~km~s$^{-1}$), suggesting that these lines are tracing the same material. However, there are some differences between the two molecular tracers. First, we observe that HCN(4-3) is much stronger in the {\it southern extension} than CS(7-6). While both molecular tracers seem equally strong along the {\it southwest lobe}, even coinciding in the distribution of the gas, stronger in the northern part of the lobe, at the location of clump N, than in the southern part, CS(7-6) is obviously weaker in the {\it southern extension}. This result would suggest that the {\it southern extension} is composed by denser but colder gas than the {\it southwest lobe}. However, we noted in the previous section that the {\it southern extension} is highly excited. Also, because of the strong detection of NH$_3$(6,6) in the northern region of the {\it southern extension}, towards Sgr~A*, we acknowledge the presence of warm material in the area. Therefore, the combination of the result drawn by the previous section and the picture presented by the weak emission of CS(7-6) in the {\it southern extension}, which traces warmer but less dense material than HCN(4-3), suggests a complicated morphology in the eastern part of the CND. The gas is denser and colder towards the south, and becomes much more diffuse and warmer heading north. This point will be discussed in detail later.

\begin{figure}
\begin{center}
\epsscale{1}
\plotone{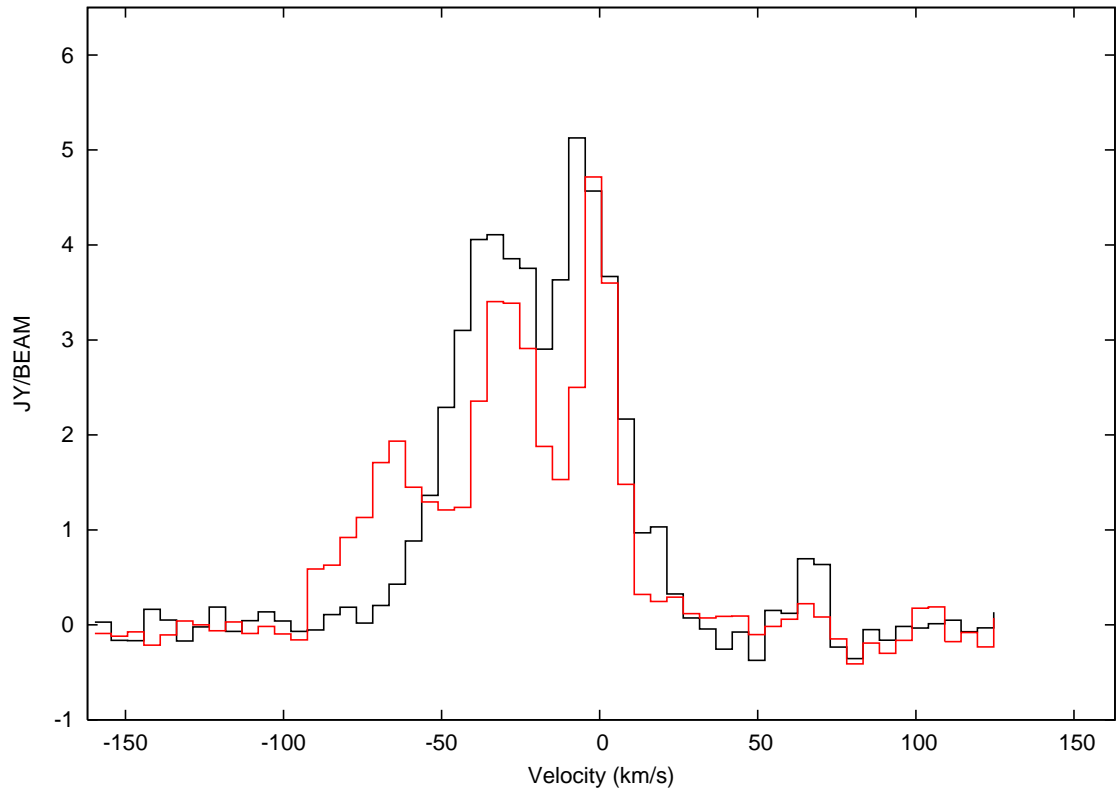}
\end{center}
\caption{Comparison of the spectra at two locations in the {\it southwest lobe} in CS(7-6). Black contours represent the spectrum in clump N in figure \ref{cs_spectra.fig}, while red contours represent the spectrum in a position 4.4$\arcsec$ east and 0.8$\arcsec$ south of clump N (location of clump N in HCN(4-3), figure \ref{spectra_hcn43.fig}).\label{clumpN.fig}}
\end{figure}

Another difference between the integrated emission from both molecular lines is that clump E in CS(7-6) does not show a double-peak profile, unlike its counterpart in HCN(4-3). The detection is much weaker, which could account for the lack of coincidence.
Spectrum N shows a two-velocity component, consistent with the result in HCN(4-3). The line profile, however, is not as wide. Clump N in the case of CS(7-6) is located a little further west than in the case of HCN(4-3). Since we have considered the interaction with the minispiral as probably affecting the profile of clump N in HCN(4-3), it could be that the greater distance from clump N in CS(7-6) to Sgr~A West and a consequently weaker interaction is responsible for the narrower profile (compare figures \ref{cs_cont.fig} and \ref{hcn43_cont.fig}). In order to check this conjecture, we extracted a spectrum closer to the inner side of the CND, 4.4$\arcsec$ east and 0.8$\arcsec$ south of clump N in CS(7-6) (at the location of clump N in HCN(4-3)) and overplotted the two spectra (figure \ref{clumpN.fig}). The broadening in the profile of the spectrum closer to the minispiral is remarkable. We conclude that the interaction with the minispiral is probably affecting the line profile. At the same time, the absorption features detected in these spectra can be caused by the negative bowls. When we examine the spectra L and M, taken nearby, we see that the lack of short-spacings may well affect the spectra towards clump N.

\begin{figure}
\begin{center}
\epsscale{1}
\plotone{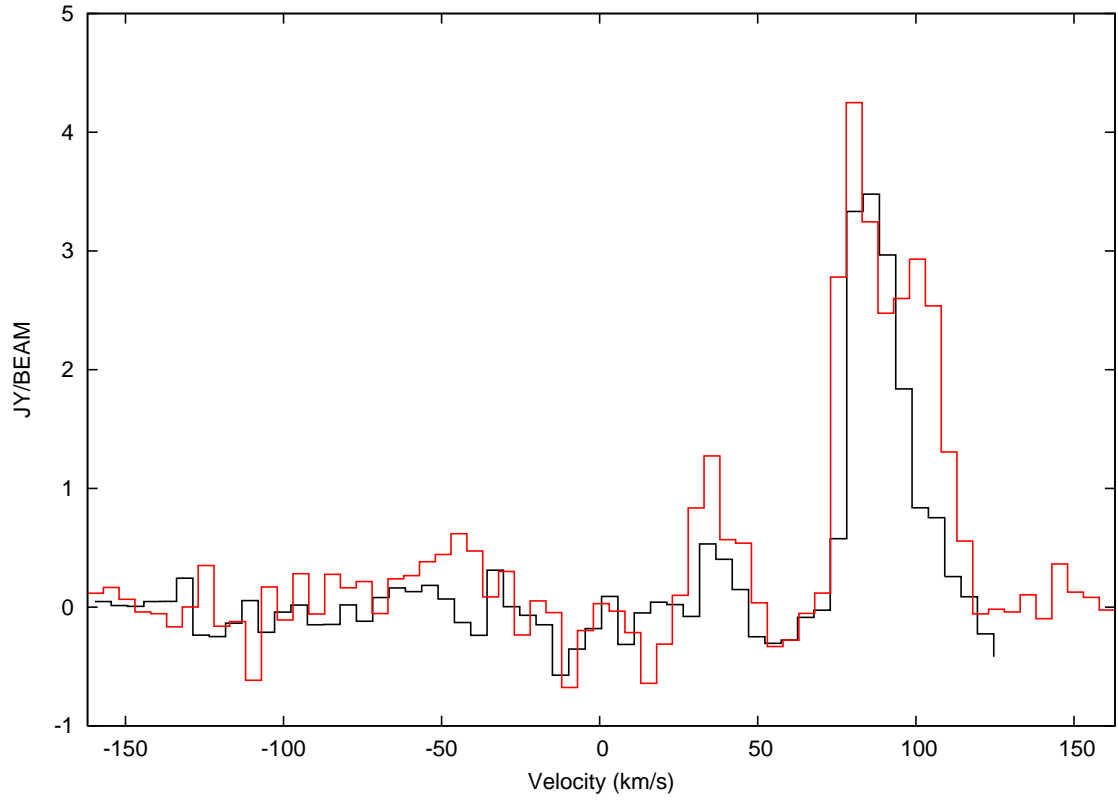}
\end{center}
\caption{Comparison of the spectra at the location of clump K in figure \ref{cs_spectra.fig}, in the {\it southwest lobe}, in CS(7-6) and HCN(4-3). Black contours represent the spectrum at the location of clump K in CS(7-6), while red contours represent the spectrum in HCN(4-3).\label{clumpK.fig}}
\end{figure}

\begin{figure}
\begin{center}
\epsscale{1}
\plotone{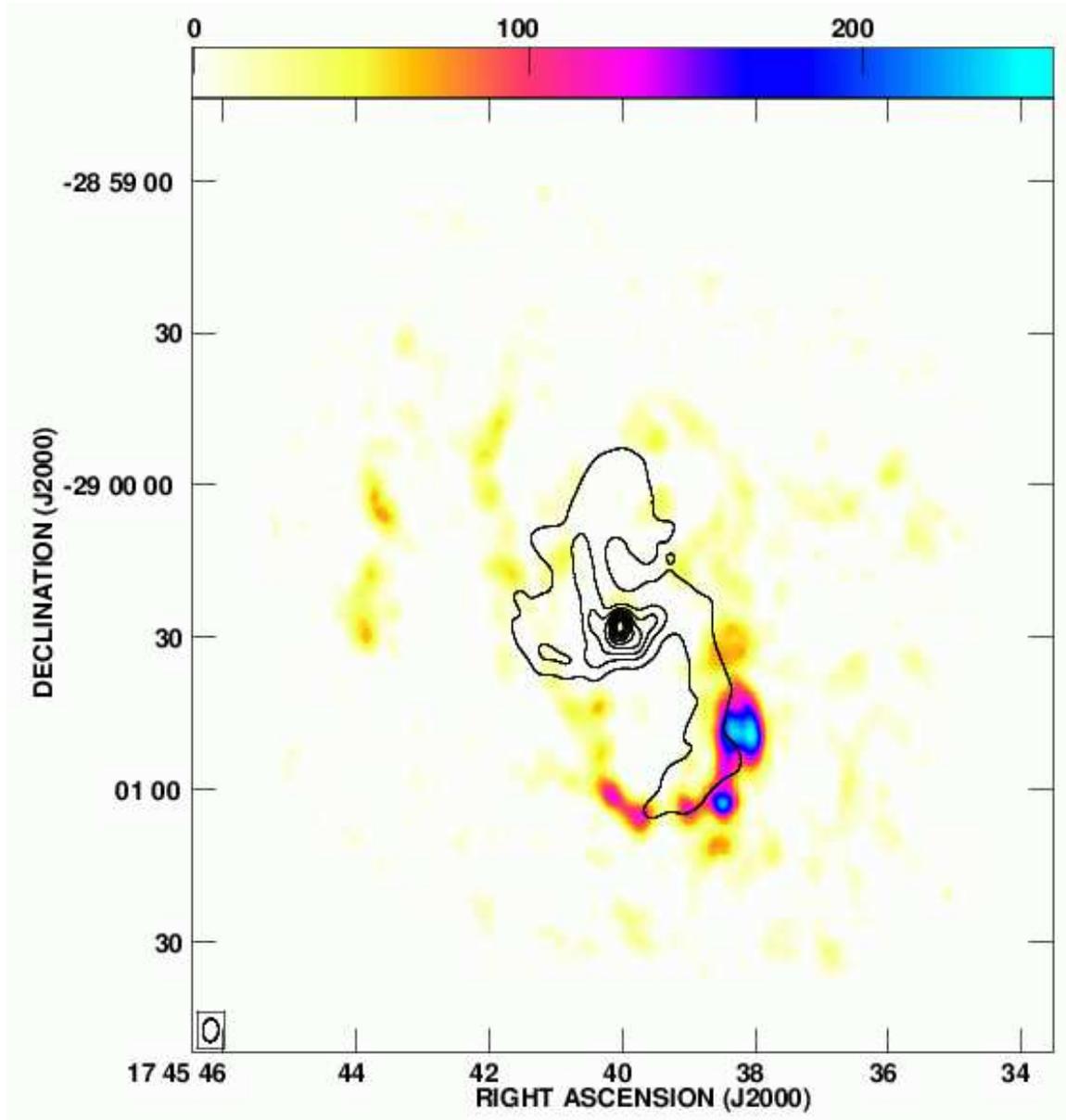}
\end{center}
\caption[Comparison map of CS(7-6) integrated intensity and 6~cm continuum integrated intensity in the Galactic center.]{6~cm continuum integrated intensity in contours (from \citep{yus87}). CS(7-6) integrated intensity in false color-scale. Contour levels are as in figure \ref{hcn43_cont.fig}. The false color-scale is in ~Jy~beam$^{-1}$~km~s$^{-1}$.\label{cs_cont.fig}}
\end{figure}

Clump K presents a very surprising profile, unlike any other. It is clearly redshifted and very narrow, more so than the ``70~km~s$^{-1}$ cloud'', which is barely detected as clump H, but the negative bowls might be seriously affecting it if we pay attention to spectrum L, taken nearby. Nonetheless, the profile is very unexpected and it is probably worth a deeper study. The comparison of the spectrum in the same position in HCN(4-3) (figure \ref{clumpK.fig}) shows a similar profile. Is once again the presence of the minispiral affecting the line profiles in the {\it southwest lobe}? Both spectra seem to be suffering from absorption around +~50~km~s$^{-1}$, a situation especially noticeable in the HCN(4-3) spectrum. However, this absorption feature is not present anywhere else in the CND.

Clump DD produces a very remarkable spectrum, with two clearly separated peaks. The spectrum at the same position in HCN(4-3) shows this profile as well. There could be some interaction involved in this location as absorption does not seem to be responsible for the ``dip'' and the negative bowls detected in the vicinity because of the lack of short-spacings are not so prominent. As we mentioned in the previous section, interaction with the extended bar of the minispiral is the most probable explanation of this broad profile.

Spectrum R shows a double peak profile and, observing the important negative bowls nearby (spectrum T) it is very probable that the short-spacings problem is the cause of this double peak profile. This problem is probably also affecting clumps FF, U and W, if we compare with spectra T and V. Therefore, as in the case of HCN(4-3), the lack of short-spacings is affecting the data. However, because of the lack of single-dish observations of CS(7-6) in the Galactic center, we can not calculate the amount of missing flux.

When we compare the CS(7-6) detected emission with the 6~cm continuum emission from \cite{yus87}, we observe that the gap in the CND in the CS(7-6) emission towards the north coincides with the position of the northern arm of the minispiral, with clumps E and F delimiting the gap to the west and clump D to the east (figure \ref{cs_cont.fig}).

\begin{figure}
\begin{center}
\plotone{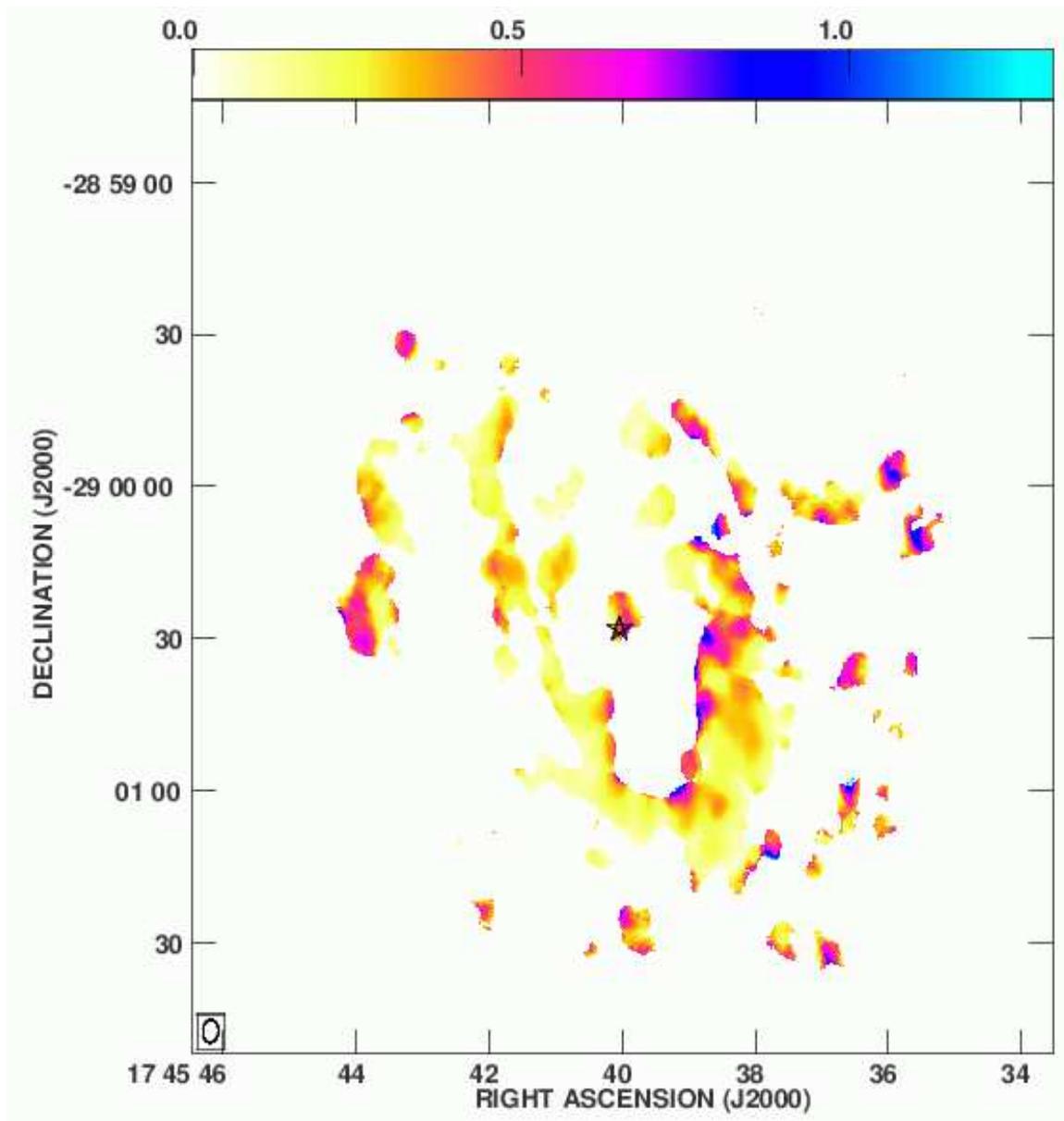}
\end{center}
\caption{Ratio of CS(7-6) and HCN(4-3) integrated intensity. Sgr A* is marked with a star.\label{cs_ratio.fig}}
\end{figure}

The ratio between CS(7-6) and HCN(4-3) (calculated using only the pixels with flux~$\ge$~3$\sigma$) is useful to better understand the distribution of the dense material (figure \ref{cs_ratio.fig}). We can observe that the ratio increases towards the inner edge of the CND, especially in the western part. Coincidentally, the western arc of the minispiral is observed in the same region (figure \ref{cs_cont.fig}). Both CS(7-6) and HCN(4-3) have similarly high critical densities, with HCN(4-3) tracing slightly denser gas than CS(7-6), but the temperature traced by CS(7-6) is double the temperature traced by HCN(4-3) (49~K vs. 25~K). We could interpret the ratio distribution as being ruled mostly by the temperature, although the density should also be a factor. Therefore the molecular gas in the inner side of the ring is probably warmer and could be less dense. As we mentioned before, the temperature increase can be due to the absorption of UV radiation emitted by the nuclear stellar cluster. High-ratio values found to the west, outside of the CND, can also be related to the presence of higher-temperature gas, as the {\it western streamer} is very clearly detected in NH$_3$(3,3) (tracing gas at $\sim$~125~K).

The comparison of CS(7-6) and NH$_3$(3,3) (figure \ref{cs7_6_33.fig}), CS(7-6) smoothed to match the NH$_3$ resolution, NH$_3$(3,3) data from \cite{mcg01}) shows that both molecular tracers coincide at the location of the strongest peak within the CND (as was also noted regarding HCN(4-3)) and the {\it northeast arm}. However, the coincidences stop there, with CS(7-6) and NH$_3$(3,3) having in general very different distributions in the region observed in both tracers. The NH$_3$(3,3) map by \cite{mcg01} is much larger than the CS(7-6) map, and it is not represented here in its entirety. The correlation between the high-ratio values that we found previously west of the CND and the {\it western streamer} cannot be checked in this figure because of the lower resolution of the NH$_3$(3,3) data, and the small clumps detected in CS(7-6) have been smoothed out.

When comparing CS(7-6) and NH$_3$(6,6) we observe that the eastern part of the CND and the {\it eastern arm} are the only regions where both molecular tracers overlap and are well correlated (figure \ref{cs7_6_66.fig}). The comparison is similar to that obtained with HCN(4-3) and NH$_3$(6,6), except for the northern part of the CND, which is barely detected in CS(7-6). Once again, we note that NH$_3$(6,6) is detected inside the central cavity, whereas CS(7-6) is limited to the CND.  This result is consistent with the idea that the CND is better traced by higher-density and lower-temperature tracers since only a minimum part of the CND is traced by NH$_3$, while CS (and HCN) are detected along most of the structure.

Similarly to the way we proceeded in the previous section, we overplot the NH$_3$(6,6) map and the CS(7-6)/HCN(4-3) ratio (CS(7-6)/HCN(4-3) ratio convolved to the NH$_3$(6,6) resolution, figure \ref{66_ratiocs.fig}). Unlike in the case of the HCN(4-3)/HCN(1-0) ratio map, the CS(7-6)/HCN(4-3) ratio does not correlate with the NH$_3$(6,6) emission. The highest ratio values, as mentioned before, are found in the western side of the CND, overlapping with the western arc of the minispiral (when higher-resolution is utilized since the NH$_3$(6,6) resolution is not sufficient to observe such details). If the CS(7-6)/HCN(4-3) ratio is tracing the region with the highest temperature-lowest density combination, then the western part of the CND is either warmer or less dense or both. Since NH$_3$(6,6) is detected more strongly in the eastern part of the CND, it indicates the presence of high-temperature gas in this region of the CND. Therefore, the western part of the CND may be less dense than the eastern part, and the high-ratio could be in this case a sign of lower density instead of higher temperature. The material detected by the high-temperature tracers might be unrelated to the material observed in high-density but lower-temperature tracers. This case is indeed true in the eastern part of the CND, as proved by the comparison of the HCN(4-3) and NH$_3$(6,6) spectra, showed previously in figure \ref{66_spec.fig}. However, the kinematic study indicates the coincidence of the material traced in the western part of the CND by both HCN(4-3) and NH$_3$(6,6). At the same time, the strength of the detection of the warmer gas is much smaller in the western part of the CND than in the eastern part. This indicates that the warmer and more diffuse gas is well mixed with the colder and denser gas in the western part, but not in the eastern part, where the amount of warmer gas is much larger, although not in the totality of the {\it southern extension} where it is only detected in the northernmost region. The lack of strong HCN(1-0) emission in the {\it southern extension}, tracing the same material as HCN(4-3) (figures \ref{hcn4_3_hcn.fig} and \ref{ratio43_10.fig}), but characterized by a lower temperature and critical density, indicate a higher temperature and density than in the {\it southwest lobe}. Furthermore, if a transition defined by a lower temperature and critical density, such as HCN(1-0), coincides in its spatial distribution in the southern part of the CND with a transition characterized by a higher-temperature but also lower critical density, CS(7-6) (both of them compared to HCN(4-3)), the {\it southern extension} will be then denser than the southwestern part of the CND. Since the material coming from the {\it southern streamer} is approaching the CND in the region of the {\it southern extension}, if the latter structure appears as denser it may be related with the possible interaction previously mentioned between the CND and the {\it southern streamer}. Because of this interaction, the material forming the southeastern part of the CND may be undergoing a compression process, appearing as denser than the adjacent region.

\section {Mass estimates}

In the previous section we noted the clumpy nature of the CND. As mentioned, \cite{jac93,shu04} and \cite{chr05} found that the different clumps (or cores) within the CND were dense enough to overcome the tidal shear produced by the central supermassive black hole, and therefore the CND might be a more stable structure than previously thought \citep{gus87}. For our results, we use the Virial calculation to estimate the required masses and densities for a clump to be stable against tidal shear. We consider then, the clumps to be gravitationally bound against the motions of the gas in the clump, with uniform density and where optical depths do not play a significant role in line-broadening \citep{roh00}.

\begin{equation}
{\it M~=~250 \left(\frac {\Delta v_{1/2}}{km~s^{-1} }\right)^{2} \left(\frac {R}{pc}\right) (M_\odot)},
\label{eqn:virialmass}
\end{equation}

where $\Delta${\it v$_{1/2}$} is the FWHM velocity linewidth and {\it R} is the radius of the clump.

The results are displayed in tables \ref{tab:hcn} and \ref{tab:cs}. Table \ref{tab:hcn} shows the masses measured for the clumps marked in figure \ref{hcn4_3.fig} and table \ref{tab:cs} displays the results for the clumps in figure \ref{cs7_6.fig}.

The virial masses calculated by \cite{chr05} range from 2~$\times$~10$^{3}$ to 89~$\times$~10$^{3}$~M$_\odot$, while the results obtained by \cite{shu04} are slightly smaller (3~$\times$~10$^{3}$ to 45~$\times$~10$^{3}$~M$_\odot$). Our measurements are similar, although a little bit higher than these previous results(4~$\times$~10$^{3}$ to 595~$\times$~10$^{3}$~M$_\odot$). We have not calculated the masses for all the clumps due to some difficult non-gaussian profiles (like clump R), that prevented us from being able to calculate a linewidth. The mean virial density of the clump, assuming spherical symmetry, can be calculated with:

\begin{equation}
{\it \overline{\rho}~=~\frac {3M} {4 \pi R^{3}}}
\label{eqn:virialdensity}
\end{equation}

Therefore, the internal virial density of the clump is:

\begin{equation}
{\it n_{H_2}~=~ \frac {3 M} { 4 \pi R^{3} m_{H_2}}}
\label{eqn:internaldensity}
\end{equation}

The critical density for a clump to survive the tidal shear from a 4~$\times$~10$^{6}$~M$_\odot$ black hole is calculated using the model from \citet{vol00}, which assumes the mass distribution in the inner region of the Milky Way to be described by a spherically symmetric approximation:

\begin{equation}
{\it M~=~4~\times~10^{6} + 1.6~\times~10^{6} (\frac{D}{pc})^{1.25} (M_\odot),}
\label{eqn:massdistribution}
\end{equation}
  
where D is the radius from the center of the Galaxy (i.e. distance to Sgr~A*).

The critical density for a clump to be tidally stable is then:

\begin{equation}
{\it n_{H_2}~=~2.87~\times~10^{7} [(\frac{D}{pc})^{-3} + 0.4 (\frac{D}{pc})^{-1.75}](cm^{-3})}
\label{eqn:criticaldensity}
\end{equation}

For a 1.5~pc distance, using equation \ref{eqn:criticaldensity}, we can calculate the critical density to be 1.4~$\times$~10$^{7}$~cm$^{-3}$. Therefore, we can conclude that the clumps we have detected using both HCN(4-3) and CS(7-6) are tidally stable, as it can be seen in tables \ref{tab:hcn} and \ref{tab:cs}.

\cite{san98} reported that the total ionized gas of the minispiral is 10$^{2}$~M$_\odot$, and \cite{jac93} measured a neutral gas mass in the northern arm of the minispiral of 3~$\times$~10$^{2}$~M$_\odot$. We can then assume the minispiral with a total mass of $\sim$~4~$\times$~10$^{2}$~M$_\odot$. At the same time, kinematic models by \cite{san98}, using an infalling gas velocity of half the rotation velocity (mentioned previously as 110~km~s$^{-1}$), were able to explain the majority of the characteristics of the minispiral. Therefore, the infalling time should be $\sim$~3~$\times$~10$^{4}$~yr. Inflow models support the idea of the minispiral as being formed by a cloud that lost angular momentum, probably due to cloud-to-cloud collisions and then fell towards the center. While infalling, the UV radiation from the nuclear stellar cluster dissociated the molecular gas and later ionized the neutral atomic gas \citep{jac93}. Consequently, if the material composing Sgr~A West comes directly from the CND, which has a total molecular mass of 1.3~$\times$~10$^{6}$~M$_\odot$ (calculated with the HCN(4-3) results), and the current mass of the minispiral takes 3~$\times$~10$^{4}$ yr to reach the center, if a mere 10\% of the CND is stripped off and becomes part of the minispiral (as \cite{chr05} considered), the lifetime of the CND is $\sim$~9~$\times$10$^{6}$~yr, much longer that its rotation period (8~$\times$~10$^{4}$~yr). Therefore, the CND is not a transient structure, since the amount of time needed for the CND to disappear, ``swallowed'' by the inner cavity, is longer than the amount of time needed to circle it. However, \cite{her02} found the molecular gas approaching Sgr~A*, as traced by NH$_3$(6,6), not to follow the minispiral path. Consequently, the previous lifetime value can be an overestimation, since not all the molecular gas approaching the dynamical center of the Milky Way is contained in the minispiral. Nonetheless, if the CND is a non-transient structure, the clumpiness should be expected to disappear and the CND to become a homogeneous structure. However, the observations do not show such homogeneity, instead, the clumps present a wide range of velocities. This dispersion in velocities could result in various interactions between the clumps (cloud-to-cloud collisions), with the more diffuse material spiraling inwards towards the center. Finally, recent studies by \cite{mar07} showed that the accretion rate of Sgr~A* is between 2~$\times$~10$^{-9}$ and 2~$\times$~10$^{-7}$~M$_\odot$ yr$^{-1}$. If the infalling rate is 1.5~$\times$~10$^{-2}$~M$_\odot$ yr$^{-1}$ (considering for now only the material confined in Sgr~A West), there is clearly a surplus of material inside the cavity that is not accreting into the black hole. The remaining material could be undergoing star formation processes. \cite{kra91} reported that 10$^{3}$ - 10$^{4}$~M$_\odot$ in the inner cavity became stars around 10$^{6}$ years ago. With the infalling rate that we are considering, an accumulation of 10$^{4}$~M$_\odot$ will take $\sim$~7~$\times$~10$^{5}$~yr (and even less if we consider the material traced by NH$_3$(6,6)). Therefore, the material infalling towards Sgr~A* could be forming stars in the inner cavity, and not only accreting into the black hole.


\section{Summary}

We have successfully detected HCN(4-3) and CS(7-6) within 2~pc of Sgr~A*, effectively tracing the CND. We demonstrate that the higher HCN transition minimizes the self-absorption problem observed in lower transitions \citep{gus87,wri01,chr05}, therefore providing a much more reliable sampling of the kinematics and structure of the CND.

The emission from both molecular tracers, HCN(4-3) and CS(7-6), appears in clumps, forming a 'necklace-like' structure around CND. The clumps have various sizes, from $\sim$~3$\arcsec$ to $\sim$~13$\arcsec$. In the case of HCN(4-3), the emission detected from these clumps amounts to 61\% of the single-dish flux detected by \cite{mar95}. Problems due to missing short spacings, however, continue to affect the observed kinematics. Nevertheless, when smoothing our data to match the resolution of the JCMT results from \cite{mar95}, we can easily note the remarkable similarity between the smoothed (to a 15$\arcsec$ resolution) HCN(4-3) emission map and the HCN(4-3) integrated emission map from \cite{mar95}. This result suggests that the extended emission does not greatly contribute to the final result and the molecular gas is mostly concentrated in clumps. Therefore, the missing short spacing problem probably does not seriously affect the final morphological results.

Moreover, we confirm the stability of the CND based on the density of the clumps, which is large enough for the clumps to overcome the tidal shear produced by the supermassive black hole. Furthermore, the lifetime value of the CND is far longer than the rotation value. This result supports the conclusions of \cite{jac93,shu04} and \cite{chr05}, agreeing that the CND is not a transient structure.

Both HCN(4-3) and CS(7-6) are found to be much more abundant in the southern part of the CND than in the northern part of the CND. In order to determine the geometrical distribution of the gas, the location of the colder and warmer gas, as well as the denser and more diffuse gas, and the implications of such distribution, line ratio measurements have been used. Our results indicate that the northern and the southern parts of the CND have different excitation levels, as well as different densities, with the southern part of the CND warmer and denser than the northern part. Also, comparing our results with those from NH$_3$, a high-density, high-temperature molecular tracer, we conclude that the molecular gas forming the CND is denser and colder than the molecular gas inside the inner cavity. More precisely, the southeastern part of the CND is denser than the southwestern part. However, the southeastern part seems to become more diffuse or warmer as the material heads northwest, approaching the supermassive black hole, as detected by NH$_3$(6,6). The comparison of the linewidths supports this conclusion, since both the NH$_3$(6,6) and HCN(4-3) present similar line profiles along the CND, kinematically tracing the same material, except in the region where the NH$_3$(6,6) emission is detected closer to Sgr~A*, where it shows a clear line-broadening effect, absent from the HCN(4-3) profiles. This result indicates a probable infall of the material forming the CND towards the dynamical center through the eastern part of the ring-like structure. However, the material detected inside the CND by the NH$_3$(6,6) is hotter and denser than the material forming the minispiral, and both structures appear to be unrelated, as indicated by \cite{her02}. At the same time, the detection of an interaction between the {\it southern streamer}, as traced by NH$_3$(3,3), and the CND indicates that the material forming the ring may be undergoing a compression process due to the material approaching the CND before the gas spirals inward towards the center. Therefore, the gas detected in the southeasternmost part of the CND may be pushed towards the northwest and heated in the process, explaining the lack of HCN detection.

In addition, we have detected a correlation between the minispiral and the CND. Line-broadening has been detected in the spectra of the clumps that spatially coincide with the arcs of the minispiral. We have observed this phenomenon in the eastern, western and northern parts of the CND. The western arc of the minispiral and the inner western part of the CND seem to overlap. Also, we have observed that the eastern end of the extended bar of the minispiral spatially coincides, at least in projection, with the location of a clump characterized by a broad spectrum. A similar situation has been observed regarding the northern part of the CND and the northern arm of the minispiral, where line-broadening has also been found in the spectra of the clumps seemingly located at the very end of the northern arm. Furthermore, we have detected a gap in the northern region of the CND, which the northern arm of Sgr~A West appears to be traversing.

These results suggest that the CND and the minispiral may be physically related. Sgr~A West could be gas which has been stripped from the CND. Alternatively, the spiral arms could be features infalling from beyond the CND. While more detailed kinematics are needed, we detect the influence of the minispiral on the western inner part, the northern part, and the eastern part of the CND, where we can see line-broadening in the vicinity of Sgr~A West.

Finally, we have observed that the inner edge of the CND seems more highly-excited than the outer part of the ring, as traced by the ratio of HCN(4-3)/HCN(1-0). Most probably the nuclear stellar cluster is responsible for the excitation of the inner side of the CND.


\acknowledgments
We thank M. Christopher for providing the HCN(1-0) data we have used for comparison. We also thank the referee, T. Wilson, for his helpful suggestions to improve the manuscript.
During the development of this study, M.M.-C. has been supported by a Smithsonian Institution Visiting Student Grant and an Academia Sinica Institute of Astronomy and Astrophysics Fellowship.



\clearpage

\begin{figure}
\begin{center}
\epsscale{1}
\plotone{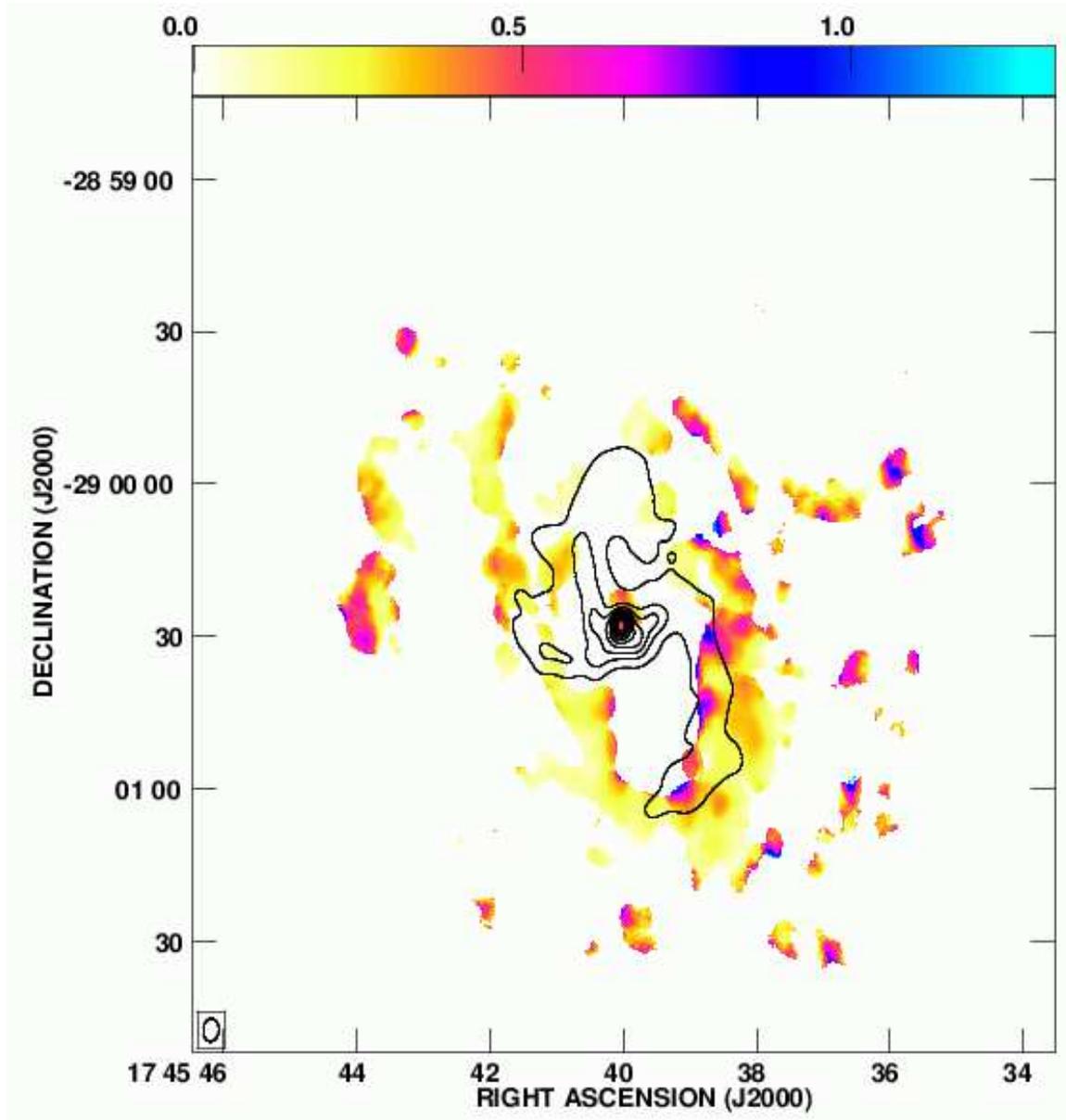}
\end{center}
\caption{6~cm continuum integrated intensity in contours (from \cite{yus87}). Ratio of CS(7-6) and HCN(4-3) in color-scale. Contour levels are as in figure \ref{hcn43_cont.fig}.\label{ratio_cont.fig}}
\end{figure}

\begin{figure}
\begin{center}
\epsscale{1}
\plotone{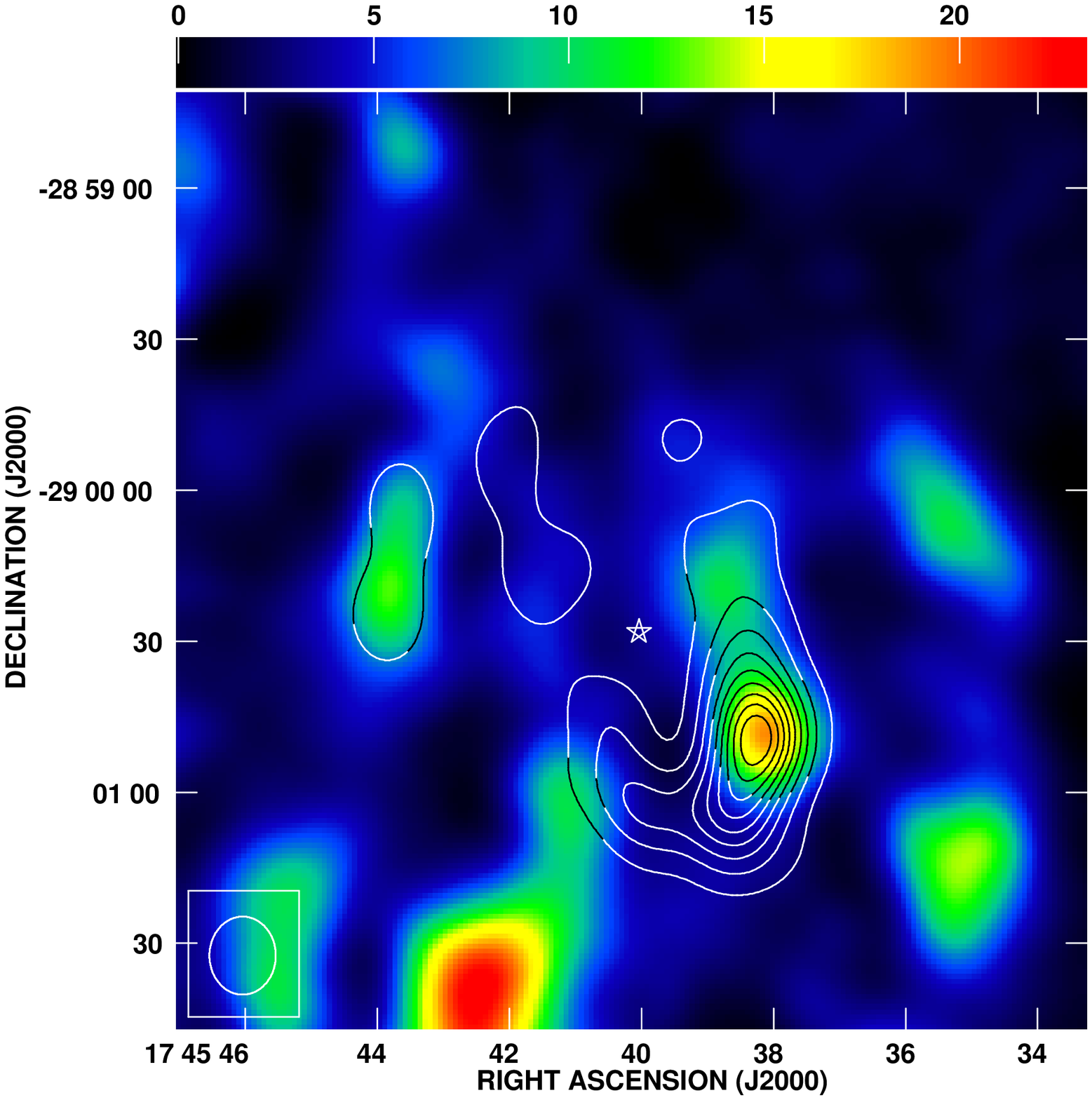}
\end{center}
\caption{CS(7-6) integrated intensity in contours. NH$_3$(3,3) integrated intensity in false color-scale from \cite{mcg01}. The CS(7-6) image has been smoothed to match the resolution of the NH$_3$(3,3) image. Contour levels are in steps of 10\% of the intensity peak, from 5.1$\times$~10$^{1}$ to 45.9$\times$~10$^{1}$~Jy~beam$^{-1}$~km~s$^{-1}$. The false color-scale is in ~Jy~beam$^{-1}$~km~s$^{-1}$. Sgr~A* is marked with a star.\label{cs7_6_33.fig}}
\end{figure}

\begin{figure}
\begin{center}
\epsscale{1}
\plotone{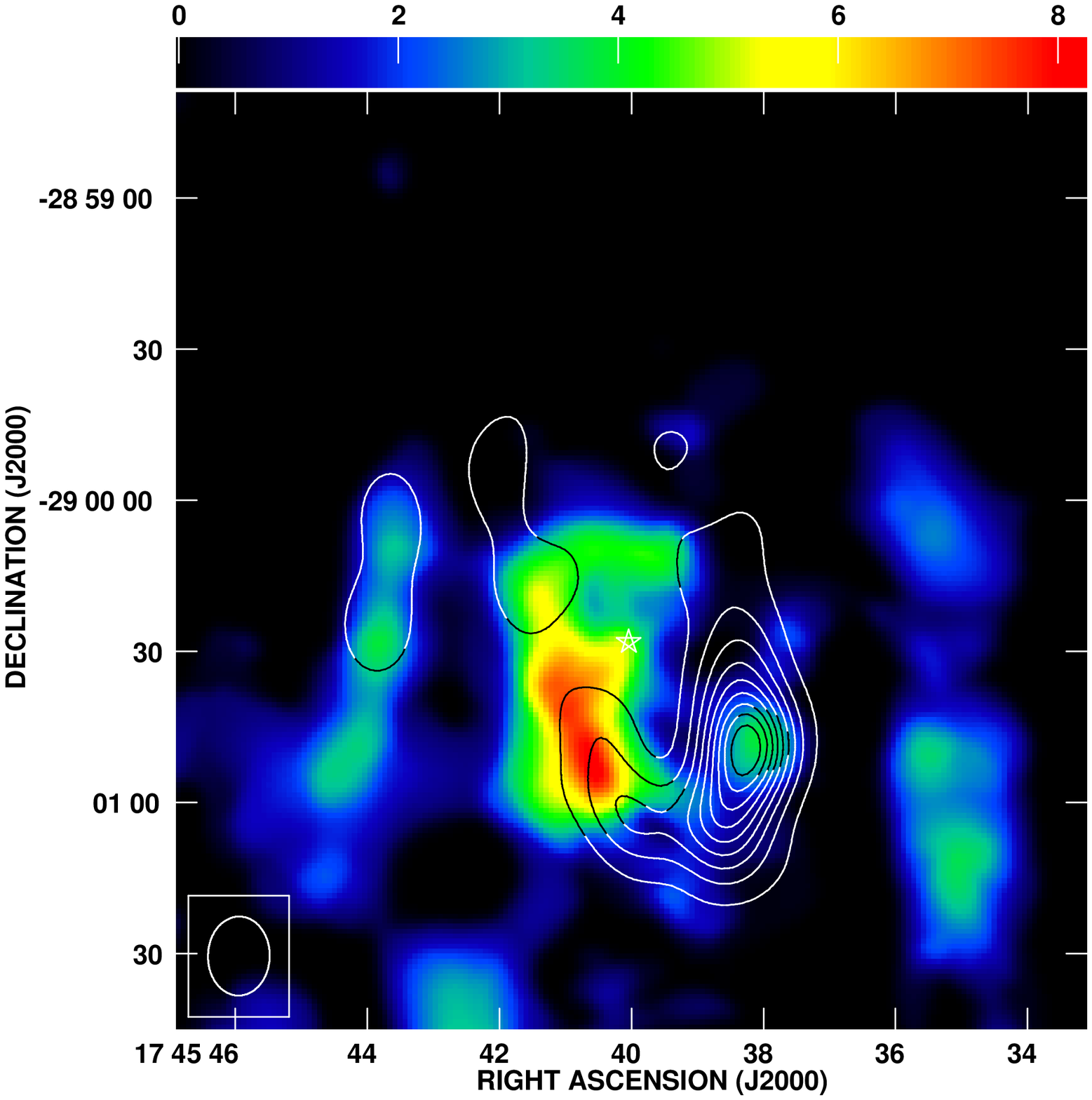}
\end{center}
\caption{CS(7-6) integrated intensity in contours. NH$_3$(6,6) integrated intensity in false color-scale from \cite{her02}. The CS(7-6) image has been smoothed to match the resolution of the NH$_3$(6,6) image. Contour levels are in steps of 10\% of the intensity peak, from 5.0$\times$~10$^{1}$ to 45.0$\times$~10$^{1}$~Jy~beam$^{-1}$~km~s$^{-1}$. The false color-scale is in ~Jy~beam$^{-1}$~km~s$^{-1}$. Sgr~A* is marked with a star.\label{cs7_6_66.fig}}
\end{figure}

\begin{figure}
\begin{center}
\epsscale{1}
\plotone{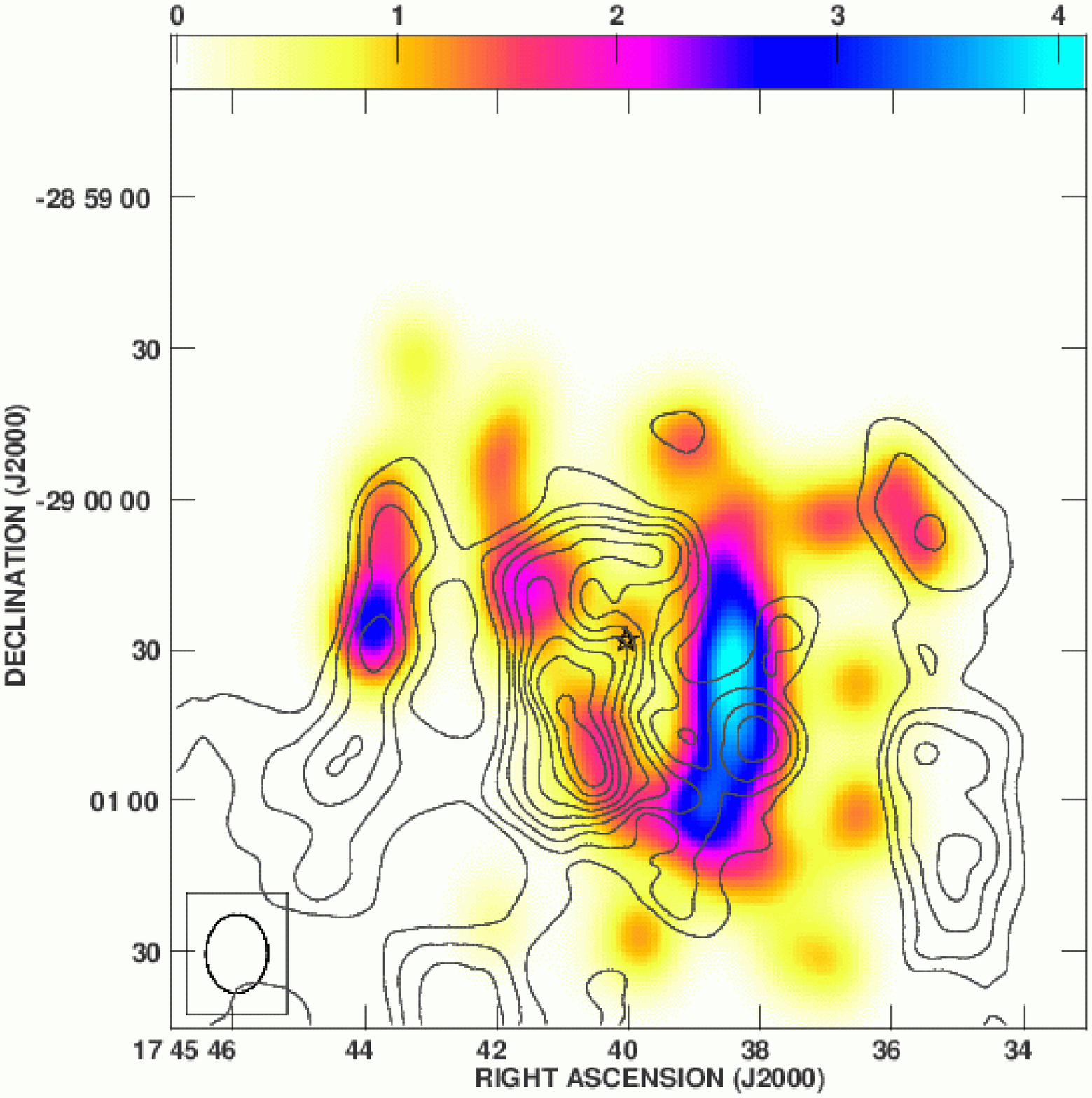}
\end{center}
\caption{Ratio of CS(7-6) and HCN(4-3) in color-scale. NH$_3$(6,6) integrated intensity in contours from \citet{her02}. Contour levels are in steps of 10\% of the intensity peak, from 8~$\times$~10$^{-1}$ to 74~$\times$~10$^{-1}$~Jy~beam$^{-1}$~km~s$^{-1}$. Sgr~A* is marked with a star.\label{66_ratiocs.fig}}
\end{figure}

\clearpage


\begin{table}
\begin{center}
\caption[Calculated masses for some of the detected peaks in HCN(4-3) in the Galactic center.] {Calculated masses for some of the detected peaks in HCN(4-3) marked in figure \ref{spectra_hcn43.fig} in the Galactic center.}
\begin{tabular}{c c c c c} 
\hline 
\hline
Clump & R & $\Delta$v & Virial Mass & Virial density  \\  
  & (pc) & (km~s$^{-1}$) & (10$^{3}$~M$_\odot$) & (10$^{7}$~cm$^{-3}$)  \\ 
\hline
{\it A} & 0.139 & 38.5 & 51.39 & 9.3 \\ 
{\it C} & 0.087 & 28.5 & 17.63 & 13.0 \\ 
{\it E} & 0.098 & 55.0 & 74.16 & 38.1 \\ 
{\it F} & 0.092 & 40.0 & 36.85 & 22.9  \\ 
{\it H} & 0.074 & 21.5 & 8.57 & 10.2  \\ 
{\it I} & 0.103 & 36.0 & 33.31 & 14.8 \\ 
{\it K} & 0.110 & 29.0 & 23.18 & 8.4 \\
{\it N} & 0.253 & 97.0 & 594.88 & 17.7 \\
{\it Q} & 0.161 & 51.0 & 104.90 & 11.7 \\
{\it U} & 0.110 & 47.0 & 60.42 & 22.2 \\
{\it W} & 0.112 & 53.5 & 7.44 & 27.4 \\
{\it X} & 0.111 & 36.5 & 37.21 & 13.0 \\
{\it Z} & 0.146 & 49.0 & 87.64 & 13.6 \\
{\it AA} & 0.145 & 49.5 & 89.02 & 2.9 \\
{\it BB} & 0.088 & 17.0 & 6.35 & 4.5 \\
{\it CC} & 0.126 & 31.5 & 31.40 & 7.5 \\
{\it EE} & 0.083 & 30.0 & 6.25 & 5.3 \\ 
\hline
\label{tab:hcn}
\end{tabular}
\end{center}
\end{table}

\begin{table}
\begin{center}
\caption[Calculated masses for some of the detected peaks in CS(7-6) in the Galactic center.]{Calculated masses for some of the detected peaks in CS(7-6) marked in figure \ref{cs_spectra.fig} in the Galactic center.}
\begin{tabular}{c c c c c} 
\hline 
\hline
Clump & R & $\Delta$v & Virial Mass & Virial density \\  
  & (pc) & (km~s$^{-1}$) & (10$^{3}$~M$_\odot$) & (10$^{7}$~cm$^{-3}$) \\ 
\hline
{\it A} & 0.066 & 23.5 & 9.16 & 15.1 \\ 
{\it D} & 0.113 & 22.5 & 14.32 & 4.8 \\ 
{\it E} & 0.080 & 31.0 & 19.34 & 17.9 \\
{\it F} & 0.093 & 29.0 & 19.51 & 11.7 \\
{\it I} & 0.083 & 30.0 & 18.66 & 15.8 \\
{\it H} & 0.064 & 15.9 & 4.09 & 7.5 \\
{\it K} & 0.151 & 20.6 & 16.09 &  2.2  \\
{\it N} & 0.183 & 52.5 & 126.05 & 9.9 \\
{\it Q} & 0.090 & 38.5 & 33.51 & 21.8 \\
{\it U} & 0.119 & 33.5 & 33.28 & 9.6 \\
{\it W} & 0.076 & 35.5 & 24.17 & 25.8 \\
{\it Z} & 0.099 & 20.0 & 9.93 & 4.9 \\
{\it BB} & 0.083 & 14.5 & 4.36 & 3.7 \\
{\it CC} & 0.135 & 21.5 & 15.58 & 3.1 \\
{\it EE} & 0.054 & 19.8 & 5.28 & 16.2\\
{\it FF} & 0.077 & 22.5 & 9.75 & 10.3 \\
\hline
\label{tab:cs}
\end{tabular}
\end{center}
\end{table}

\end{document}